\documentclass[journal]{IEEEtran}

 \usepackage{amsmath,array}

\usepackage{multirow}
\usepackage[english]{babel}
\usepackage{float}
\usepackage{breqn}
\usepackage{footnote}
\usepackage[caption=false,font=footnotesize]{subfig}
\usepackage{balance}
\usepackage{blindtext, graphicx}
\usepackage{booktabs}
\usepackage{soul,color}
\usepackage{multirow}
\usepackage{bm}
\usepackage{amsmath}
\usepackage{array}
\usepackage{pifont}
\usepackage{algorithm}
\usepackage[capitalise]{cleveref}
\usepackage{multirow}
\usepackage{algpseudocode}
\usepackage{algorithm}
\ifCLASSINFOpdf
 
\else
\fi

\begin{document}

\title{PANDA: \underline{P}rocessing-in-MRAM \underline{A}ccelerated\\ De Bruij\underline{n} Graph based \underline{D}NA \underline{A}ssembly}
\author{\IEEEauthorblockN{
Shaahin Angizi\IEEEauthorrefmark{2},
Naima Ahmed Fahmi\IEEEauthorrefmark{1},
Wei Zhang\IEEEauthorrefmark{1} and
Deliang Fan\IEEEauthorrefmark{2}}

\IEEEauthorblockA{\IEEEauthorrefmark{2}School of Electrical, Computer and Energy Engineering, Arizona State University, Tempe, AZ 85287\\ \IEEEauthorrefmark{1}Department of Computer Science, University of Central Florida, Orlando, FL 32816\\
sangizi@asu.edu, fnaima@knights.ucf.edu, wzhang.cs@ucf.edu, dfan@asu.edu }}
\maketitle

\begin{abstract}

Spurred by widening gap between data processing speed and data communication speed in Von-Neumann computing architectures, some bioinformatic applications have harnessed the computational power of Processing-in-Memory (PIM) platforms. However, the performance of PIMs unavoidably diminishes when dealing with such complex applications seeking bulk bit-wise comparison or addition operations. 
In this work, we present an efficient Processing-in-MRAM Accelerated De Bruijn Graph based DNA Assembly platform named \textit{PANDA} based on an optimized and hardware-friendly genome assembly algorithm. \textit{PANDA} is able to assemble large-scale DNA sequence data-set from all-pair overlaps. 
We first design \textit{PANDA} platform that exploits MRAM as a computational memory and converts it to a potent processing unit for genome assembly. \textit{PANDA} can execute not only efficient bulk bit-wise X(N)OR-based comparison/addition operations heavily required for the genome assembly task but a full-set of 2-/3-input logic operations inside MRAM chip.
We then develop a highly parallel and step-by-step hardware-friendly DNA assembly algorithm for \textit{PANDA} that only requires the developed in-memory logic operations. The platform is then configured with a novel data partitioning and mapping technique that provides local storage and processing to fully utilize the algorithm-level's parallelism.
The cross-layer simulation results demonstrate that \textit{PANDA} reduces the run time and power, respectively, by a factor of 18 and 11 compared with CPU. Besides, speed-ups of up-to 2-4$\times$ can be obtained over recent processing-in-MRAM platforms to perform the same task. 

\end{abstract}


\IEEEpeerreviewmaketitle

\section{Introduction}

\begin{figure}[t]
\begin{center}
\begin{tabular}{c}
\includegraphics [width=1\linewidth]{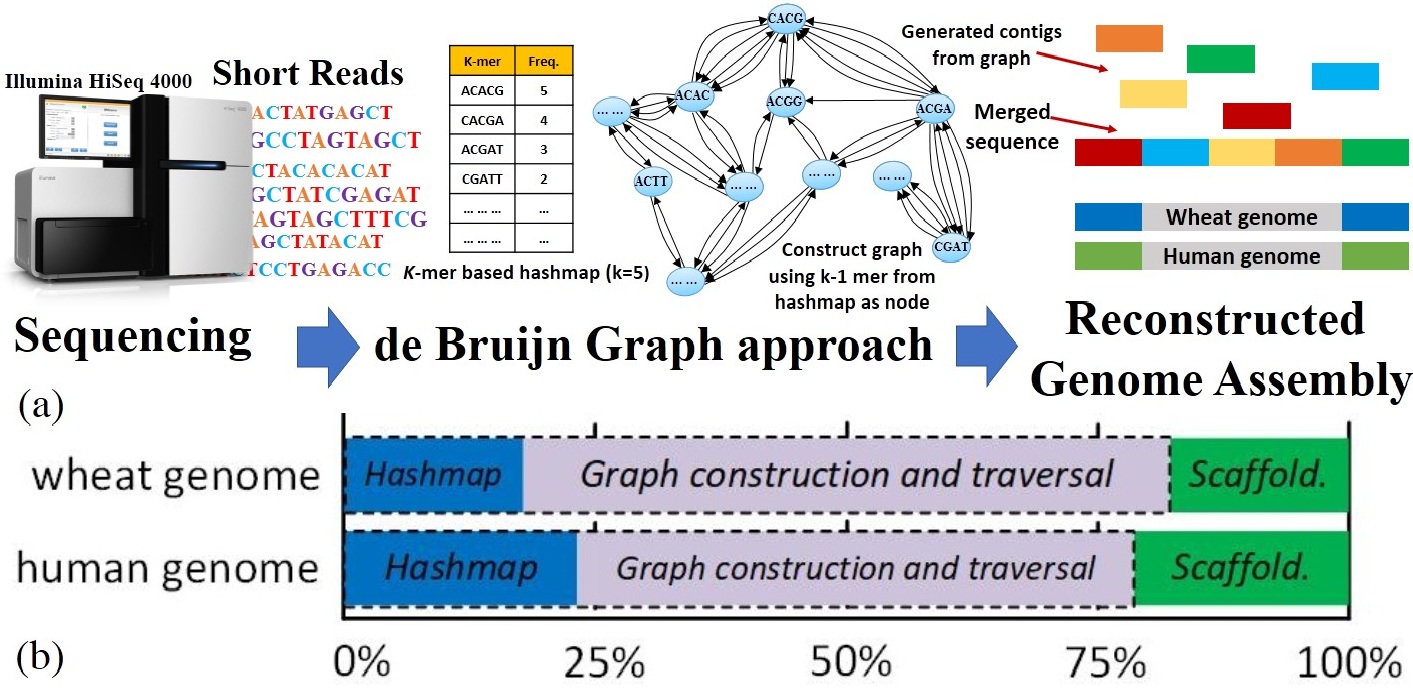}\\
\end{tabular} \vspace{-1.1em}
\caption{(a) The de Bruijn graph-based genome assembly process, (b) Break down of execution time of Meraculous genome assembler for human and wheat data-set \cite{chapman2011meraculous,georganas2014parallel}.}
\label{exe}
\end{center}\vspace{-2.1em}
\end{figure}

With the advent of high-throughput second generation parallel sequencing technologies, the process of generating fast and accurate large-scale genomics data has become a significant advancement. Such data can enable us to measure the molecular activities in cells more accurately by analyzing the genomics activities, including mRNA quantification, genetic variants detection, and differential gene expression analysis. Thus, by understanding the transcriptomic diversity, we can improve phenotype predictions and provide more accurate disease diagnostics \cite{li2010survey}. However, the reconstruction of the full-length transcripts considering sequencing errors is a challenging task in terms of computation and time. Since the current cDNA sequencing technology cannot read whole genomes in one step \cite{georganas2014parallel}, the data produced by the sequencer is extensively fragmented due to the presence of repeated chunks of sequences, duplicated reads, and large gaps. Thus, the goal of genome assembly process is to combine these large number of fragmented short reads and merge them into long contiguous pieces of sequence (i.e. contigs), to reconstruct the original chromosome from which the DNA is originated as shown in Fig. \ref{exe}a. 


Today's bioinformatics application acceleration solutions are mostly based on the von-Neumann architecture with separate computing and memory components connecting via buses and inevitably consumes a large amount of energy in data movement between them \cite{zokaee2018aligner,angizi2019aligns}. In the last two decades, Processing-in-Memory (PIM) architecture, as a potentially viable way to solve the memory wall challenge, has been well explored for different applications \cite{angizi2019aligns,pourmeidani2020probabilistic,seshadri2017ambit,angizi2019graphide,yu2017skeleton,angizi2020pim,angizi2018cmp}. Especially processing-in-non-volatile memory architecture has achieved remarkable success by dramatically reducing data transfer energy and latency \cite{li2016pinatubo,chowdhury2020dna,kang2017memory,angizi2019graphs,roohi2019apgan}. The key concept behind PIM is to realize logic computation within memory to process data by leveraging the inherent parallel computing mechanism and exploiting large internal memory bandwidth. Besides, most of CPU \cite{li2009soap2}-/ GPU \cite{liu2012soap3}-/ FPGA \cite{arram2016leveraging}- and even PIM \cite{zokaee2018aligner,angizi2019aligns}-based efforts have only focused on the DNA short read alignment problem, while the de novo genome assembly problem still relies mostly on CPU-based solutions \cite{mahmood2011gpu}. De novo assemblers are categorized into Overlap Layout Consensus (OLC), greedy, and de Bruijn graph-based designs. Recently, de Bruijn graph-based assemblers have gained much more attention as they are able to solve the problem using Euler path in a polynomial time rather than finding Hamiltonian path in OLC-based assemblers as an NP hard problem \cite{varma2016fpga}. There are multiple CPU-based genome assemblers implementing the bi-directed de Bruijn graph model, such as Velvet \cite{zerbino2008velvet}, Trinity \cite{grabherr2011full}, etc. However, only a few GPU-accelerated assemblers have been presented such as GPU-Euler \cite{mahmood2011gpu,goswami2018gpu,ren2018efficient}. This mainly comes from the nature of the assembly workload that is not only compute-intensive but also extremely data-intensive requiring very large working memories. Therefore adapting such problem to use GPUs with their limited memory capacities has brought many challenges \cite{lu2013gpu}. A graph-based genome assembly process, shown in Fig. \ref{exe}a, as the main focus of this work, basically consists of multiple stages, i.e. \textit{k}-mer analysis for creating a Hashmap, graph construction and traversal, and scaffolding and gap closing. Fig. \ref{exe}b depicts the breakdown of execution time for the well-known Meraculous assembler \cite{chapman2011meraculous} for the human and wheat data sets. We observe that Hashmap and graph construction/ traversal are the two most expensive components, which together take over 80\%  of the total run time.

This motivates us to show that the genome assembly problem and especially computationally-loaded components can exploit the large internal bandwidth of Magnetic Random Access Memory (MRAM) chip for PIM acceleration. Moreover, with a careful observation of genome assembly workload, it turns out this task heavily relies on comparison and addition operations. However, due to the intrinsic complexity of X(N)OR logic, the throughput of processing-in-memory platforms \cite{li2016pinatubo,chowdhury2020dna,angizi2019aligns,Angizi2020Assem,angizi2018pima} unavoidably diminishes when dealing with such bulk bit-wise operations. This is because multi-cycle majority/AND/OR-based  operations. In this work, we explore a highly-parallel and PIM-friendly implementation of de Bruijn graph-based genome assembly that can accelerate especially the first two stages of the algorithm. Overall this paper makes the following contributions:

(1) To the best of our knowledge, this work is the first that designs a high-throughput comparison/addition-friendly processing-in-MRAM architecture for the de Bruijn graph-based genome assembly. We develop \textit{PANDA} based on a set of innovative microarchitectural and circuit-level schemes to realize a data-parallel computational core for genome assembly; 
(2) We reconstruct the existing genome assembly algorithm in a step-by-step fashion to be fully implemented in PIM platforms. It supports short read analysis, graph construction, and traversal; 
(3) We propose a dense data mapping and partitioning scheme to process the indices locally and handle various length DNA sequences; 
(4) We extensively assess and compare \textit{PANDA}'s performance, energy-efficiency, and memory bottleneck ratio with a CPU and recent potential PIM platforms.

\section{PANDA Platform}

\subsection{SOT-MRAM}

\begin{figure}[t]
\begin{center}
\begin{tabular}{l}
\includegraphics [width=0.65\linewidth]{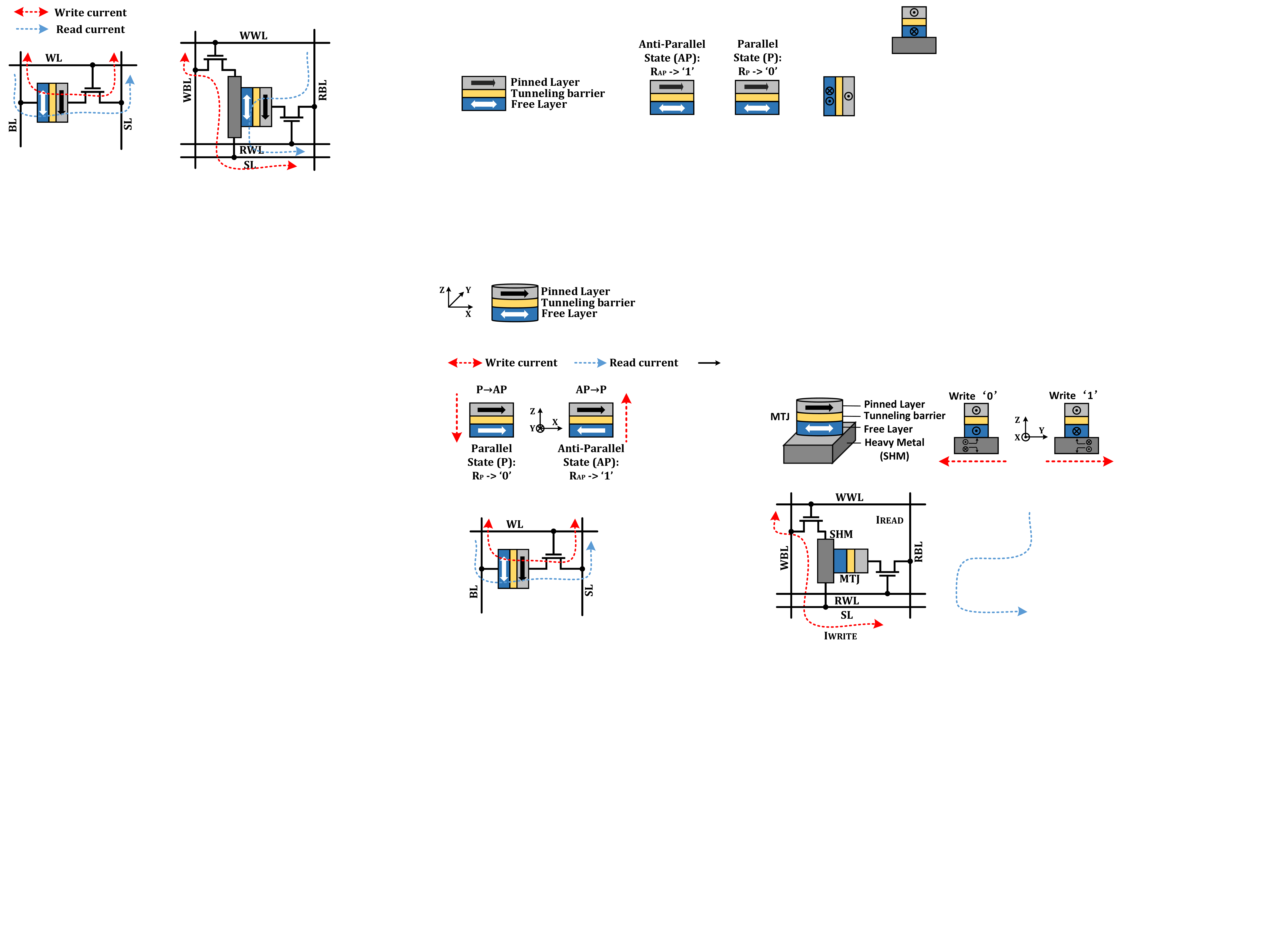}\\
 \hspace{3.5cm}   \small (a) 
 \end{tabular}
 \begin{tabular}{ll}
 \begin{minipage}[c]{0.2\textwidth}
\includegraphics [width=1\linewidth]{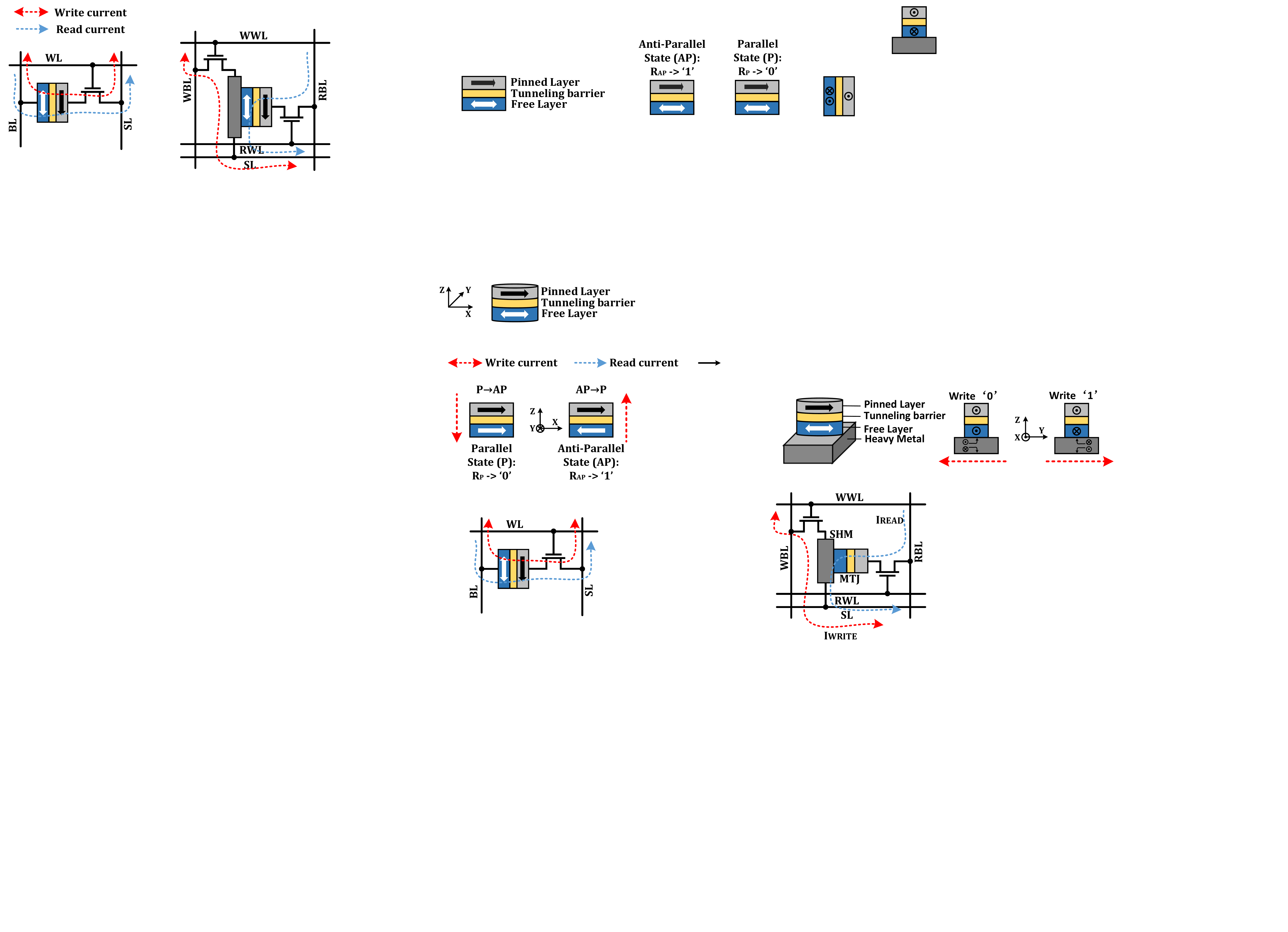} \end{minipage} & 
 \begin{minipage}[c]{0.2\textwidth}
 \scalebox{0.7}{
\begin{tabular}[t]{|c|c|c|}
\hline
Operations & \begin{tabular}[c]{@{}c@{}}Write \\ `1'(`0')\end{tabular} & Read   \\ \hline
WWL        & $V_{DD}$                                                    & 0      \\ \hline
RWL        & 0                                                         & $V_{DD}$       \\ \hline
RBL        & 0                                                         & $I_{READ}$ \\ \hline
WBL        & $V_{WP}$($V_{WN}$)                                            & 0      \\ \hline
SL         & 0                                                         & 0      \\ \hline
\end{tabular}}
 \end{minipage}
 \\
 \hspace{2.1 cm}   \small (b) & \hspace{2.1 cm}  \small (c)\\\vspace{-0.5em}
 \end{tabular} \vspace{-1em}
\caption{(a) SOT-MRAM device structure and Spin Hall Effect, (b) Schematic and (c) biasing conditions of SOT-MRAM bit-cell.} 
\label{SOTMRAM}
\end{center}
\end{figure}

Fig. \ref{SOTMRAM}a shows a Spin-Orbit Torque Magnetic Random Access Memory (SOT-MRAM) device structure. The storage element in SOT-MRAM is SHE-MTJ \cite{fong2015spin,angizi2018imce}, a composite device structure of a Spin Hall Metal (SHM) and Magnetic Tunnel Junction (MTJ). The binary data is stored as resistance states of MTJ. Data-`0'(/`1') is encoded as the MTJ's lower(/higher) resistance or parallel(/anti-parallel) magnetization in both magnetic layers (free and fixed layers). Here the flow of charge current ($\pm$y) through the SHM (Tungsten, $\beta-W$ \cite{pai2012spin}) will cause accumulation of opposite directed spin on both surfaces of SHM due to spin Hall effect \cite{fong2015spin}. Thus, a spin current flowing in $\pm$z is generated and further produces spin-orbit torque (SOT) on the adjacent free magnetic layer, causing switch of magnetization. Each cell located in the computational sub-array is connected with a Write Word Line (WWL), Write Bit Line (WBL), Read Word Line (RWL) Read Bit Line (RBL), and Source Line (SL). The bit-cell structure of 2T1R SOT-MRAM and its biasing conditions are shown in Fig. \ref{SOTMRAM}b and \ref{SOTMRAM}c, respectively.
In this work, the magnetization dynamics of Free Layer (${m}$) are modeled by LLG equation with spin-transfer torque terms, which can be mathematically described as \cite{fong2015spin}: \vspace{-1em}

\begin{multline}
\frac{d{m}}{dt}=-|\gamma|{m}\times {H_{eff}}+\alpha\bigg({m}\times \frac{d{m}}{dt}\bigg) \\ + |\gamma|\beta({m}\times {m}_{p} \times {m}) -|\gamma|\beta \epsilon' ({m} \times {m}_{p})
\end{multline}

\begin{equation}
\label{eq_beta}
\beta = |\frac{\hbar}{2\mu_0 e}| \frac{I_{c} P}{A_{MTJ} t_{FL} M_s}
\end{equation}
where $\hbar$ is the reduced plank constant, $\gamma$ is the gyromagnetic ratio, $I_{c}$ is the charge current flowing through MTJ, $t_{FL}$ is the thickness of free layer, $ \epsilon' $ is the second Spin transfer torque coefficient, and ${H_{eff}}$ is the effective magnetic field, $P$ is the effective polarization factor, $A_{MTJ}$ is the cross sectional area of MTJ, ${m_{p}}$ is the unit polarization direction. 
Note that the ferromagnets in MTJ have In-plane Magnetic Anisotropy (IMA) in x-axis \cite{fong2015spin}. With the given thickness (1.2nm) of the tunneling layer (MgO), the Tunnel Magneto-Resistance (TMR) of the MTJ is $\sim$ 171.2\%.

\subsection{Architecture Design}

\begin{figure}[t]
\begin{center}
\begin{tabular}{c}
\includegraphics [width=1\linewidth]{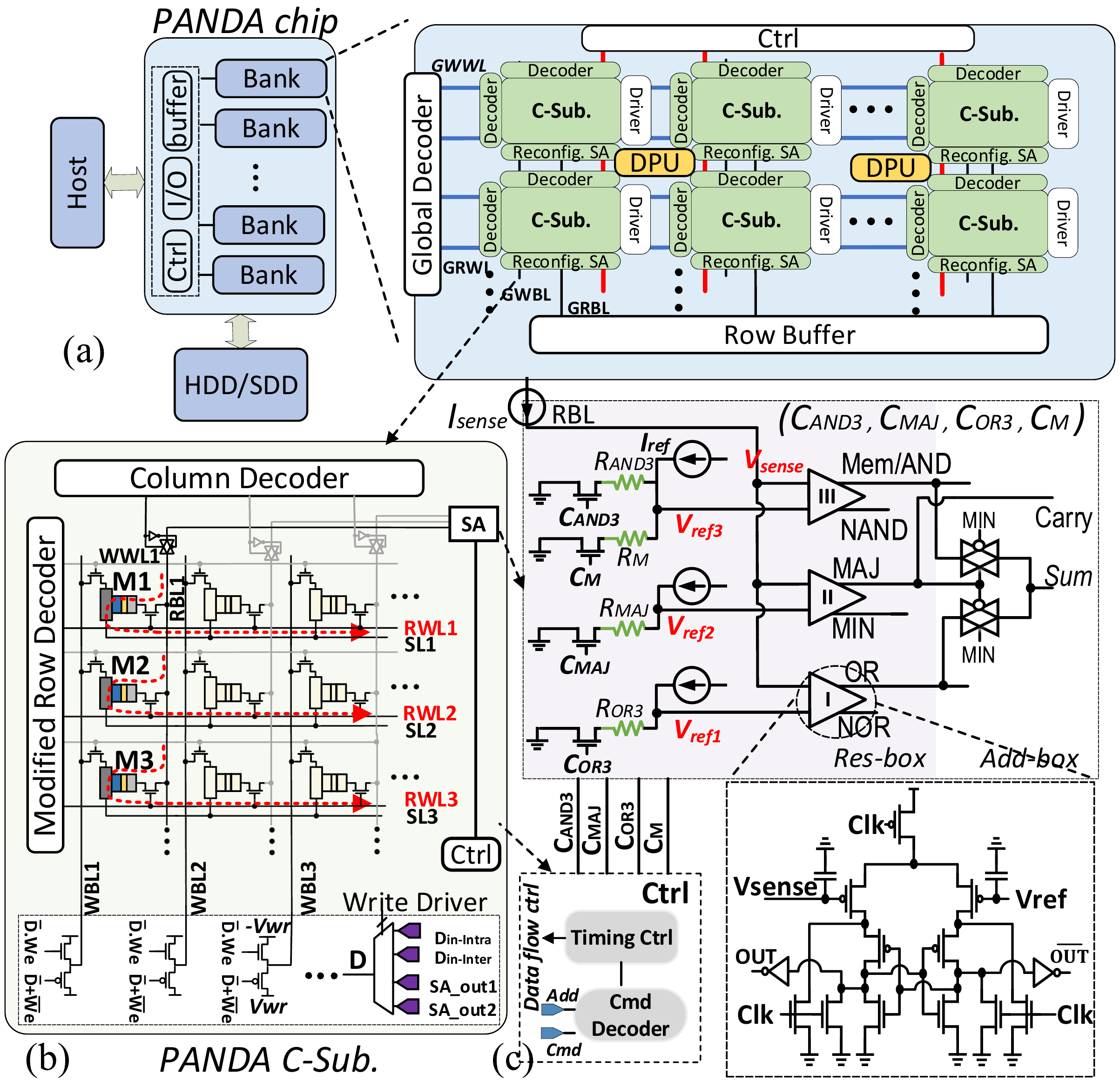}\\
\end{tabular} \vspace{-1em}
\caption{\textit{PANDA} platform: (a) Memory organization, (b) Computational sub-array, (c) The new reconfigurable sense amplifier designed to implement a full-set of 2- and 3-input logic operations.}
\label{arc_sub-array}
\end{center}\vspace{-1em}
\end{figure}

We develop \textit{PANDA} platform based on typical SOT-MRAM hierarchy. Each memory chip consists of multiple memory banks divided into 2D sub-arrays of SOT-MRAM cells as shown in Fig. \ref{arc_sub-array}a. We then apply our modification on the sub-array level to make it reconfigurable to support both memory operation and in-memory bit-line computation. As depicted in Fig. \ref{arc_sub-array}b, the computational memory sub-array (C-Sub.) of \textit{PANDA} consists of a modified memory row decoder, column decoder, write driver, and reconfigurable Sense Amplifier (SA). The data-parallel intra-sub-array computation of sub-array is timed and controlled using a Controller (ctrl) w.r.t. the physical address of operands. 

\textit{PANDA} is especially designed to support bulk bit-wise operations between operands stored in each BL. Therefore, the in-memory computational throughput is solely limited by the physical memory row size i.e. 4KB/8KB in modern main memory chips. Digital Processing Units (DPU) are also shared between computational sub-arrays to handle nonparallel computational load of the platform. In the following, we explain different elements and the supported functions by \textit{PANDA}.

\subsection{PIM Operations}

\textit{\textbf{Write Operation:}} To \textit{write} `0' (/`1') in a cell, e.g. in the cell of 1st column and 2nd row (M2 in Fig. \ref{arc_sub-array}b), the associated write driver first pulls WBL1 to negative (/positive) write voltage. This will provide a preset charge current flow from $-V_{wr}$ to GND (/$+V_{wr}$ to GND) that eventually changes the cell's resistance to Low-$R_P$/ (High-$R_{AP}$).

\begin{table}[b]
\caption{Control bits for reconfigurable SA.}\vspace{-1em}
\scalebox{0.82}{
\begin{tabular}{|c|c|c|c|c|c|c|}
\hline
Operations      & $C_{AND3}$ & $C_{MAJ}$ & $C_{OR3}$ & $C_{M}$ & Active SA   & Row Init. \\ \hline
Read            & 0          & 0         & 0         & 1       & SA-III        & No        \\ \hline
(N)AND3/(N)AND2 & 1          & 0         & 0         & 0       & SA-III      & No/Yes    \\ \hline
(N)OR3/(N)OR2   & 0          & 0         & 1         & 0       & SA-I        & No/Yes    \\ \hline
X(N)OR2         & 1          & 1         & 1         & 0       & SA-I-II-III & Yes       \\ \hline
Maj (Carry)/Min     & 0          & 1         & 0         & 0       & SA-II       & No        \\ \hline
XOR3 (Sum)      & 1          & 1         & 1         & 0       & SA-I-II-III & No        \\ \hline
\end{tabular}}
\label{enable}
\end{table}

\begin{figure}[h]
\begin{center}
\begin{tabular}{cc}
\includegraphics [width=0.41\linewidth, height=2.4 cm]{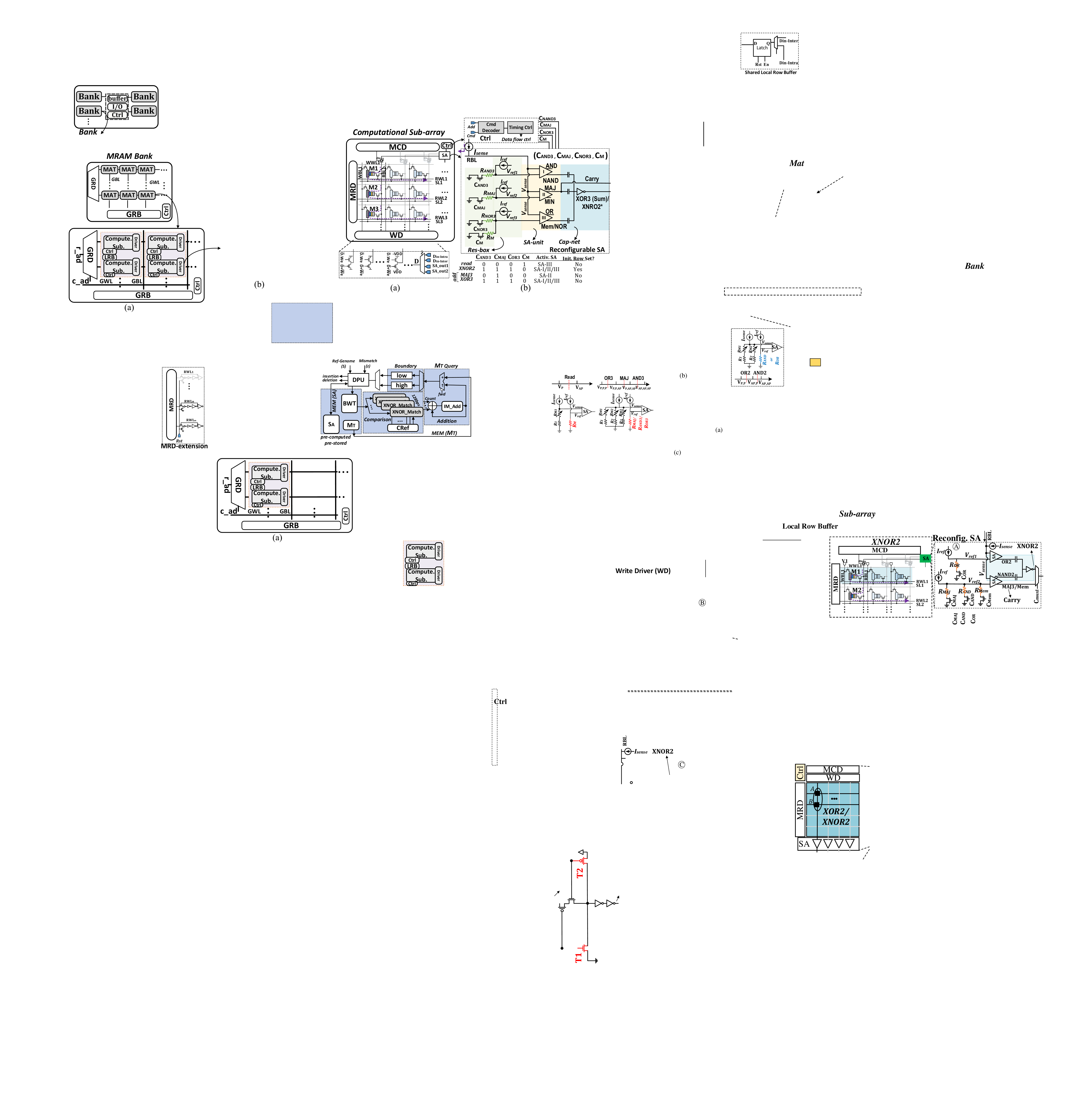}&\includegraphics [width=0.55\linewidth, height=3 cm]{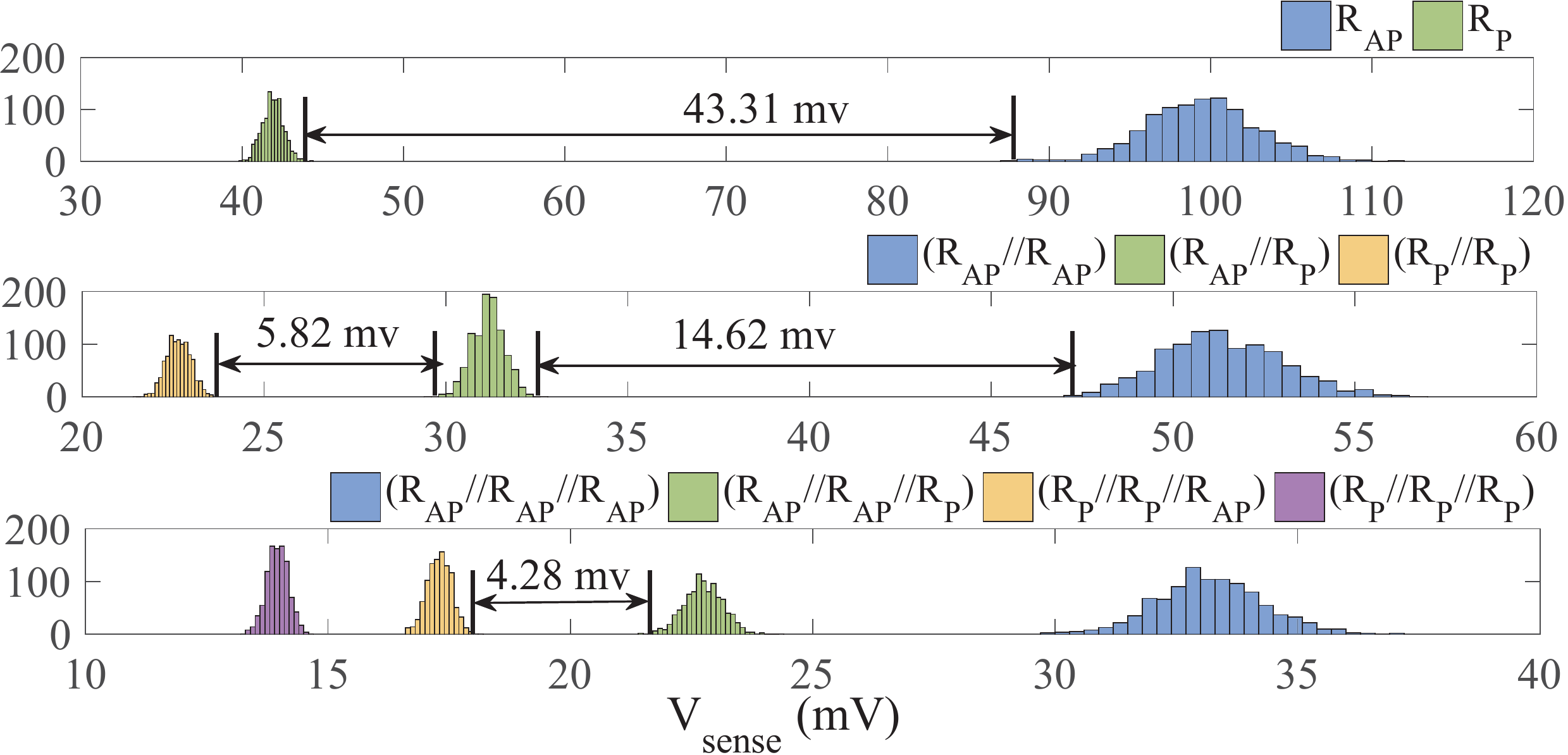}  \\ 
\small (a)  &\small (b)
 \end{tabular}\vspace{-0.8em}
\caption{(a) Reference comparison to realize in-memory operations, (b) Monte-Carlo simulation of V\textsubscript{sense}.}
\label{SARef}\vspace{-1.5em}
\end{center}
\end{figure}

\textit{\textbf{Reference Selection and Bit-line Computing:}}
\textit{PANDA} leverages the reference selection and bit-line computing method on top of a novel reconfigurable SA design shown in Fig. \ref{arc_sub-array}c to handle memory read and in-memory computation. The main idea of reference selection is to simultaneously compare the resistance state of selected SOT-MRAM cell(s) with one or multiple reference resistors in SA(s) to generate the results. \textit{PANDA}'s SA consists of three sub-SAs with a total of four reference resistors. The ctrl unit could pick the proper reference using enable control bits ($C_{AND3}$, $C_{MAJ}$, $C_{OR3}$, $C_{M}$) to realize the memory read and a full-set of 2- and 3-input logic functions, as tabulated in the Table \ref{enable}. We designed and tuned the sense circuit based on StrongARM latch \cite{razavi2015strongarm} shown in Fig. \ref{arc_sub-array}c. Each read/in-memory computing operation requires two clock phases: pre-charge (Clk `high') and sensing (Clk`low'). For instance, to realize the \textit{read} operation, the memory row decoder first activates the corresponding RWL, then a small sense current (I\textsubscript{sense}) flows from the selected cell to ground, and generates a sense voltage (V\textsubscript{sense}) at the input of SA-III. This voltage is accordingly compared with the memory mode reference voltage activated by $C_{M}$ (V\textsubscript{sense,P}$<$V\textsubscript{ref,M}$<$V\textsubscript{sense,AP}), as shown in Fig. \ref{SARef}a. The SA-III produces high (/low) voltage if the path resistance is higher (/lower) than $R_{M}$ (memory reference resistance), i.e. $R_{AP}$ (/$R_{P}$).
\textit{PANDA} could implement one-threshold in-memory operations ({\tt (N)AND}, {\tt (N)OR}, etc.) by activating multiple RWLs simultaneously, and only by activating one SA's enable at a time e.g. by setting $C_{AND3}$ to `1', 3-input {\tt AND}/{\tt NAND} logic can be readily implemented between operands located in the same bit-line. To implement 2-input logics, two rows  initialized by `0'/`1' are considered in every sub-array such that functions can be made out of 3-input functions.



\textit{\textbf{Addition:}}
\textit{PANDA}'s SA is enhanced with a unique circuit design that allows single-cycle implementation of  addition/subtraction ({\tt add/sub}) operation quite efficiently. By activating three memory rows at the same time (RWL1, RWL2, and RWL3 in Fig. \ref{arc_sub-array}b), {\tt OR3}, Majority ({\tt MAJ}) and {\tt AND3} functions can be readily realized through SA-I, SA-II, and SA-III, respectively. Each SA compares the equivalent resistance of  parallel connected input cells and their cascaded access transistors with a programmable references by SA $(R_{OR3}/R_{MAJ}/R_{AND3})$. The idea of voltage comparison between V\textsubscript{sense} and V\textsubscript{ref} to realize these functions is depicted on Fig. \ref{SARef}a. 
While there are several addition-in-memory designs in non-volatile memory domain, they typically apply a large circuitry after SA to realize a multi-cycle design. In order to implement a single-cycle addition operation, we then reformulate the full-adder Boolean expression to make it PIM-friendly. We noticed when majority function of three input is 0, the Sum can be implemented by {\tt OR3} function and when majority function is 1, Sum can be achieved through  {\tt AND3} function. This behavior can be implemented by a multiplexer circuit shown in Add-box in Fig. \ref{arc_sub-array}c. The Boolean logic of such in-memory addition function is written as: 
\begin{equation}\label{e4} 
     \small \begin{split}
     Carry & = AB+AC+BC =
      Maj(A,B,C)
     \end{split}
\end{equation} 
\vspace{-1em}
\begin{equation}\label{e5} 
     \footnotesize \begin{split}
     Sum & = ((\overline{AB+AC+BC}).(A+B+C)) + ((AB+AC+BC).(ABC))\\
     &= \overline{Maj(A,B,C)}.OR(A,B,C)+ MAJ(A,B,C).AND(A,B,C) \\
     &= \overline{Carry}.OR(A,B,C)+ Carry.AND(A,B,C) \\
     \end{split}
\end{equation} 

The carry-out of the full-adder can be directly produced by {\tt MAJ} function (Carry in Fig. \ref{arc_sub-array}c) just by setting $C_{MAJ}$ to `1' in a single memory cycle. For {\tt MAJ} operation, $R_{MAJ}$ is set at the midpoint of $R_{P}//R_{P}//R_{AP}$ (`0',`0',`1') and $R_{P}//R_{AP}//R_{AP}$ (`0',`1',`1'),  as depicted in Fig. \ref{SARef}a.
Now, assume  M1, M2, and M3 operands (Fig. \ref{arc_sub-array}b), the  \textit{PANDA} can generate Carry-{\tt MAJ} and Sum-{\tt XOR3} in-memory logics in a single memory cycle. The ctrl's configuration for such {\tt add} operation is tabulated in Table \ref{enable}.

\textit{\textbf{Comparison:}}
\textit{PANDA} platform offers a single-cycle implementation of {\tt XOR3} in-memory logic (Sum). To realize the bulk bit-wise comparison operation based on {\tt XNOR2}, one memory row in each \textit{PANDA}'s sub-array is initialized to `1'. In this way, {\tt XNOR2} can be readily implemented out of {\tt XOR3} function. Therefore, every memory sub-array can potentially perform parallel comparison operation without need to external add-on logic or multi-cycle operation.\vspace{-1em}

\subsection{Performance Analysis}

\begin{figure}[t]
\begin{center}
\begin{tabular}{c}
\includegraphics [width=0.95\linewidth]{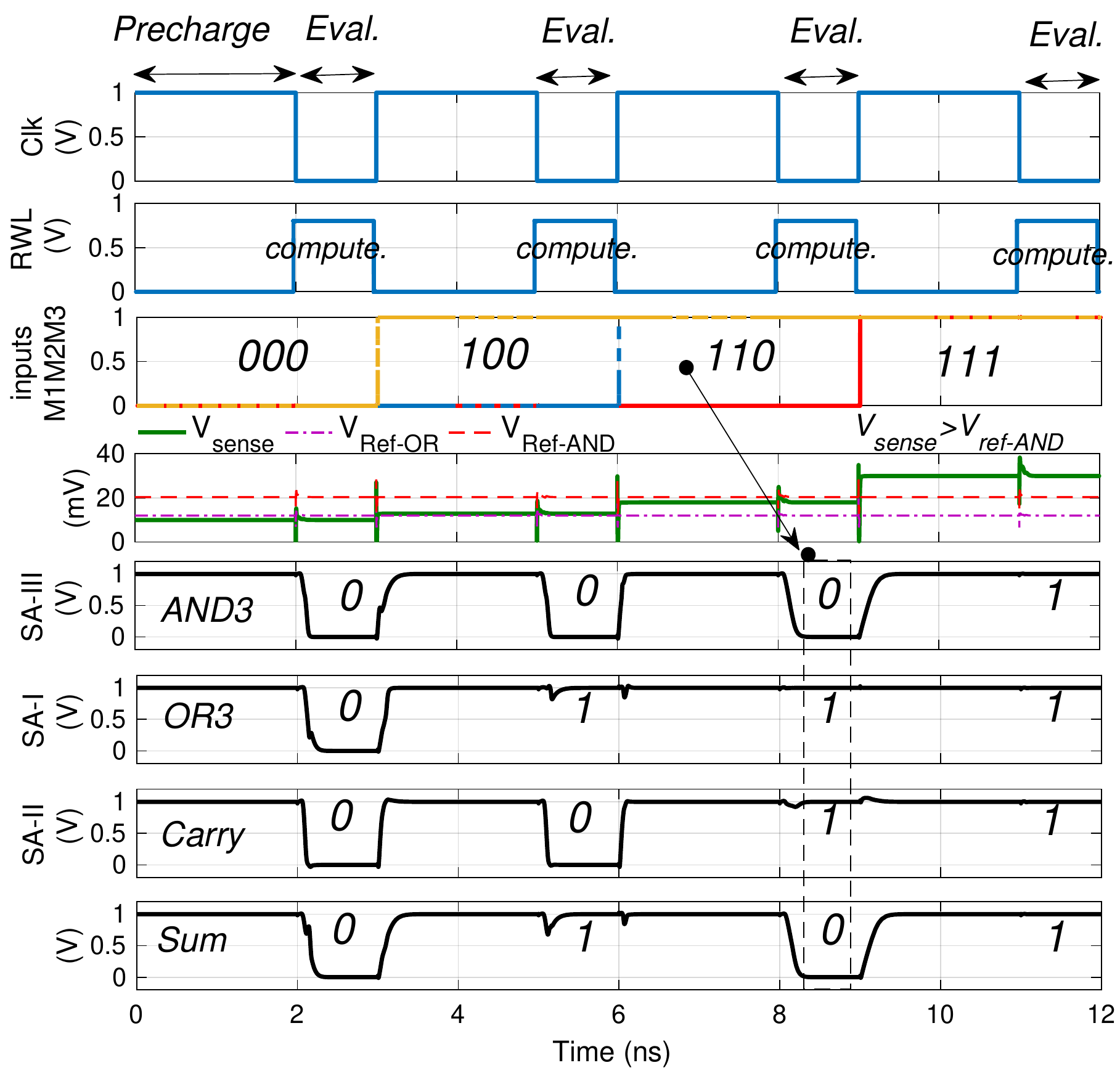}\\
\end{tabular} \vspace{-1.7em}
\caption{Transient simulation wave-forms of \textit{PANDA}'s sub-array and its reconfigurable SA for performing single-cycle in-memory operations.}
\label{wave}
\end{center}\vspace{-2.em}
\end{figure}

\textit{\textbf{Functionality:}}
To verify the circuit functionality of \textit{PANDA}'s sub-array, we first model SOT-MRAM cell by jointly applying the Non-Equilibrium Green's Function (NEGF) and Landau-Lifshitz-Gilbert (LLG) with spin Hall effect equations \cite{fong2015spin,angizi2019aligns}. We then develop a Verilog-A model of 2-transistor 1-resistor SOT-MRAM device with parameters listed in Table \ref{device} to co-simulate with other peripheral CMOS circuits displayed in Fig. \ref{arc_sub-array} in Cadence Spectre and SPICE. We use 45nm North Carolina State University (NCSU) Product Development Kit (PDK) library \cite{NCSU_PDK} for our circuit analysis. The transient simulation result of a single 256$\times$256 sub-array is shown in Fig. \ref{wave}.
\begin{table}[b]
 \caption{Device Parameters} \vspace{-0.5em} 
\centering 
\begin{tabular}{lc}
\hline
\textbf{Parameter}                              & \textbf{Value}                             \\ \hline
Free layer dimension $(W\times L\times t)_{FL}$ & $60 \times 40 \times 2$ $nm^3$             \\
SHM dimension                                   & $60 \times 80 \times 2$ $nm^3$             \\
Demagnetization Factor, $D_x$; $D_y$; $D_z$     & 0.066; 0.911; 0.022                        \\
Spin flip length, $\lambda_{sh}$                   & 1.4 $nm$                                   \\
Spin hall angle, $\theta_{sh}$                  & 0.3                                        \\
Gilbert Damping Factor, $\alpha$                & 0.007                                      \\
Saturation Magnetization, $M_s$                 & 850 $kA/m$                                 \\
Oxide thickness, $t_{ox}$                       & 1.2 $nm$                                   \\
RA product, $RA_{p}$ / $TMR$                    & $10.58$ $\Omega \cdot \mu m^2$ / $171.2\%$ \\
Supply voltage                                  & 1 $V$                                      \\
CMOS technology                                 & 45 $nm$                                    \\
SOT-MRAM cell area                              & 69 $F^2$                                   \\
Access transistor width                         & 4.5$F$                                     \\
Cell aspect Ratio                               & 1.91                                       \\ \hline
\end{tabular}
\label{device}
\end{table}
We take M1, M2, and M3 as three SOT-MRAM cells located in the first column as the inputs for our evaluation. Here, we consider four input combination scenarios for the  write operation, as indicated by 000, 100, 110, and 111 in Fig. \ref{wave}.
For the sake of clarity of wave-forms, we assume a 3ns period clock synchronises the write and read operation. However, a 2ns period can be used for a reliable read and in-memory computation.

During the precharge phase of SA (Clk=1), $\pm$V\textsubscript{write} voltage is applied to the WBL to change the MRAM cell resistance to R\textsubscript{low}=5.6$k\Omega$ or R\textsubscript{high}=15.17 $k\Omega$.  Prior to the evaluation phase (Eval.) of SA, WWL and WBL is grounded while RBL is fed by the very small sense current, I\textsubscript{sense}= 3 $\mu A$. In the evaluation phase, RWL goes high and depending on the resistance state of parallel bit-cells and accordingly SL, V\textsubscript{sense} is generated at the first input of SAs, when V\textsubscript{ref} is generated at the second input of SAs. The voltage comparison between V\textsubscript{sense} and V\textsubscript{ref} for {\tt AND3} and {\tt OR3} and the output of SAs are plotted in Fig. \ref{wave}. For example, we observe only when V\textsubscript{sense}$>$V\textsubscript{ref,AND} (M1M2M3= 111), the SA-III outputs binary `1', whereas output is `0. Fig. \ref{wave} also shows the in-memory {\tt XOR3} function (Sum) accomplished in a single memory cycle through three SA outputs.

\textit{\textbf{Reliability:}}
We assess the variation tolerance in the proposed sub-array and SA circuit by running a rigorous Mont-Carlo simulation. We run the simulation for 10000 iterations considering two source of variations in SOT-MRAM cells, first  $\sigma=5\%$ process variation on the Tunneling MagnetoResistive (TMR) and second a $\sigma =2\%$ variation on the Resistance-Area product (RA\textsubscript{P}). The results illustrated in Fig. \ref{SARef}b proves that the sense margin reduces by increasing the number of selected input cells for in-memory operations. This can be alleviated by increasing the oxide thickness $t_{ox}$ of SHE-MTJ as thoroughly discussed in \cite{angizi2019mrima}. In this way, the $t_{ox}$ was increased from 1.5nm to 2nm. This increased the sense margin by $\sim$45mV which considerably enhances the reliability.

\textit{\textbf{Sub-array level Performance:}}
To explore the hardware overhead of \textit{PANDA} on top of an standard unmodified SOT-MRAM platform, we perform an iso-capacity performance comparison. We develop both platforms with a sample 32Mb-single Bank, 512-bit Data Width in NVSim memory evaluation tool. The circuit level data is adopted from our circuit level simulation and then fed into an NVSim-compatible PIM library to report the results. Table \ref{perform} lists the performance measures for dynamic energy, latency, leakage power, and area. We observe there is a $\sim$30\% increase in the area to support the proposed in-memory computing functions for genome assembly. As for dynamic energy, the \textit{PANDA} shows an increase in R (Read) energy in spite of power gating mechanism used in the reconfigurable SA to turn off non-selected SAs (SA-I and -II while reading operation). In this way, C-Add (C stands for Computation) requires $\sim$2.4$\times$ more power compared with a single SA read operation. However, Table \ref{perform} shows \textit{PANDA} is able to offer a close-to-read latency for C-AND3 and C-Add compared with the standard design. There is also an increase in leakage power obviously coming from the add-on CMOS circuitry.
\vspace{-0.6em}

\begin{table}[h]
\caption{Performance comparison between an standard SOT-MRAM chip and \textit{PANDA}.}\vspace{-0.8em}
\scalebox{0.65}{
\begin{tabular}{|c|c|l|c|c|c|c|c|c|c|c|c|l|}
\hline
\multirow{2}{*}{Designs} & \multicolumn{2}{c|}{\multirow{2}{*}{\begin{tabular}[c]{@{}c@{}}area\\  ($mm^2$)\end{tabular}}} & \multicolumn{4}{c|}{\begin{tabular}[c]{@{}c@{}}dynamic energy \\ (nJ)\end{tabular}} & \multicolumn{4}{c|}{\begin{tabular}[c]{@{}c@{}}latency\\   (ns)\end{tabular}} & \multicolumn{2}{c|}{\multirow{2}{*}{\begin{tabular}[c]{@{}c@{}}leak. power\\ (mW)\end{tabular}}} \\ \cline{4-11}
                         & \multicolumn{2}{c|}{}                                                                          & R                  & W                  & C-AND3               & C-Add              & R                 & W                & C-AND3             & C-Add             & \multicolumn{2}{c|}{}                                                                           \\ \hline
Standard                 & \multicolumn{2}{c|}{7.06}                                                                      & 0.57               & 0.66               & -                    & -                  & 3.85              & 4.5              & -                  & -                 & \multicolumn{2}{c|}{402}                                                                        \\ \hline
PANDA                    & \multicolumn{2}{c|}{9.3}                                                                       & 0.78               & 0.69               & 0.85                 & 1.93               & 3.91              & 4.59             & 3.91               & 3.91              & \multicolumn{2}{c|}{586}                                                                        \\ \hline
\end{tabular}}\vspace{-1.8em}
\label{perform}
\end{table}

\subsection{Software Support}
\textit{PANDA} is designed to be an efficient and independent accelerator for DNA  assembly, nevertheless it needs to be exposed to programmers and system-level libraries to use it. \textit{PANDA} could be directly connected to the memory bus or through PCI-Express lanes as a third party accelerator. Thus, it could be integrated similar to that of GPUs. So, an ISA and a virtual machine for parallel and general-purpose  thread execution need to be developed like the NVIDIA's PTX \cite{GPU}. With that, at install time, the programs are translated to the \textit{PANDA}'s ISA discussed here to implement the in-memory functions listed in Table 1. We introduce \textit{PANDA\_Mem\_insert} (des, src, size) instruction to read a source data from the memory and write it back to a destination memory location consecutively. The size of input vectors for in-memory computation could be at most a multiple of \textit{PANDA}'s sub-array row size. \textit{PANDA\_Cmp} (src1, src2, size) performs parallel bulk bit-wise comparison operation between source vector 1 and 2. \textit{PANDA\_Add} (src1, src2, size) runs element-wise addition between cells located in a same column as will be explained in next section.

\vspace{-0.5em}
\section{PANDA Algorithm and Mapping}
\begin{figure}[h]
\begin{center}
\begin{tabular}{c}
\includegraphics [width=1\linewidth]{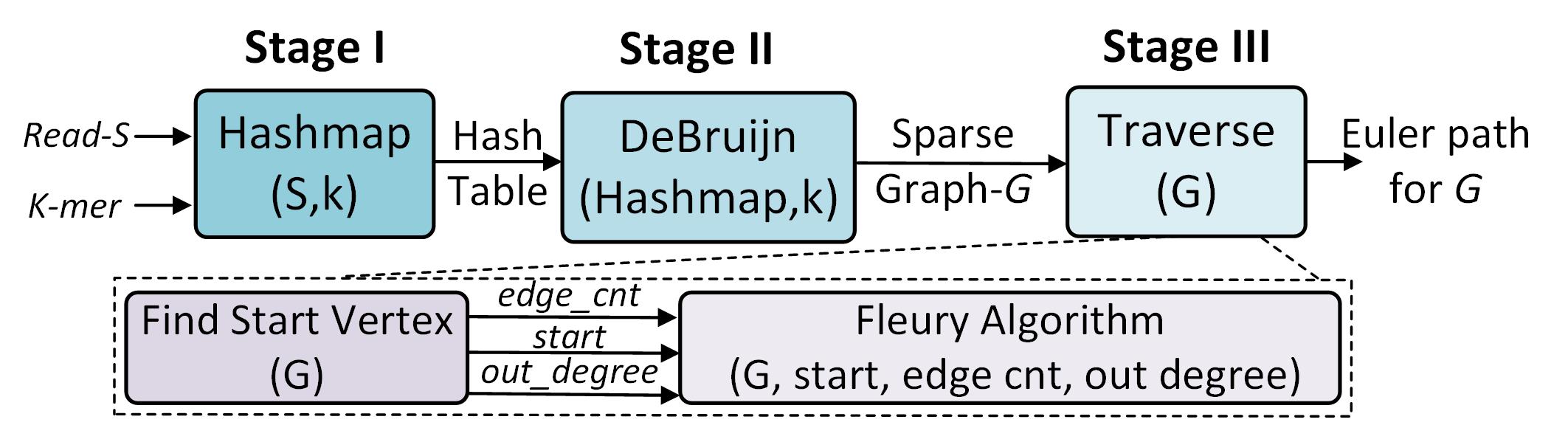}\\
\end{tabular} \vspace{-0.8em}
\caption{The genome assembly stages.}
\label{diag}
\end{center}\vspace{-1em}
\end{figure}

The genome assembly algorithm consists of three main stages visualized in Fig. \ref{diag}. First, creating a hash table out of chopped short reads (\textit{k}-mers) and keeping a count of each distinct \textit{k}-mer; second, constructing a de Bruijn Graph with Hashmap; third, traversing through de Bruijn Graph for Euler Path\footnote{The stage II and III are so-called contig. generation}. There is a final stage called scaffolding to close the gaps between contigs, which is the result of the denovo assembly \cite{georganas2014parallel}.

The first three stages always take most fraction of execute time and computational resources (over 80\%) in both CPU and GPU implementations \cite{georganas2014parallel}. To effectively handle the huge number of short reads, we modularized the assembly algorithm by focusing on parallelizing the main steps by loading only the necessary data at each stage into \textit{PANDA} platform, and leave stage-4 as our future work.

\subsection{Stage One: Hash Table}
\begin{figure}[h]
\begin{center}
\begin{tabular}{c}
\includegraphics [width=0.95\linewidth]{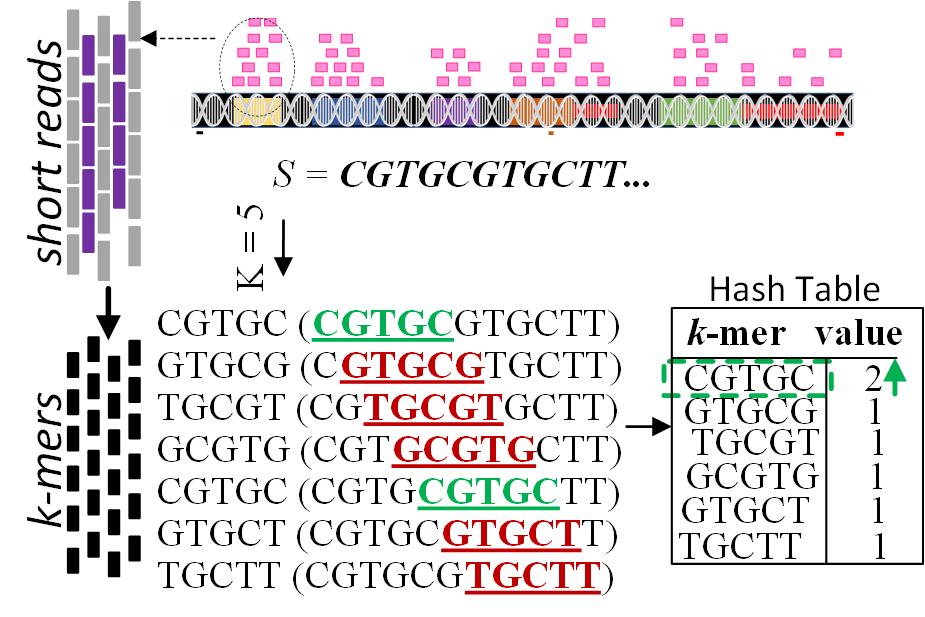}\\
\end{tabular} \vspace{-0.8em}
\caption{The hash table generation out of k-mers.}
\label{diag1}
\end{center}\vspace{-1.5em}
\end{figure}

Algorithm 1 shows the reconstructed \textit{Hashmap(S,k)} procedure in which the algorithm takes \textit{k}-mer from the original sequence (\textit{S}) in each iteration, creates a hash table entry (key) for that, and assigns its frequency (value) to 1. This step is visualized in Fig. \ref{diag1}. If the \textit{k}-mer is already in the table, it will calculate a new frequency (New\_frq) by adding the previous frequency by one and update the value. As indicated, Hashmap procedure can be implemented through \textit{PANDA\_Cmp} (comparison), \textit{PANDA\_Add} (addition), and \textit{PANDA\_Mem\_insert} (memory W/R) in-memory operations. Such functions are iteratively used in every step of ‘for’ loop and \textit{PANDA} is specially designed to handle such computation-intensive load through performing comparison, summing, and copying operations.

Considering the fact that the number of different keys in Hash table is almost comparable to the genome size $G$, the memory space requirement to save the hash is given by $\sim 2 \times G\times (k+1)$ bits (The factor of 2 is given to represent 2 bits per nucleotide). For instance, storing Hash table for human genome with $G\sim$3$\times10^9$ and $k$=32 requires $\sim$23GB mostly associated with storing the key. Due to very large memory space requirement of hash table for assembly-in-memory algorithm \cite{georganas2014parallel}, we partition these tables into multiple sub-arrays to fully leverage \textit{PANDA}'s parallelism, and to maximize computation throughput. Obviously, larger memory units \cite{li2010novo} and distributed memory schemes \cite{georganas2014parallel,simpson2009abyss} are preferable.

\begin{algorithm}[h]
  \caption{Procedure Hashmap(S, k)}
  \label{alg:algorithm1}
  \scalebox{0.67}{
    \begin{minipage}{1.5\linewidth}
  \begin{algorithmic}[1]
\Statex Step-1. Initialization:
    \State hashtable named Hashmap = \{\} 
\Statex Step-2. Fill out the table:
     \For{$i$ := 0 to length(S)-k+1}
         \State $k\_mer \gets S[i : i+k] $ \Comment{copy values of $S[i$ to $i+k]$ into variable $k\_mer$}
         \If{$\textbf{\textit{PANDA_Cmp}} (k\_mer,Hashmap) == 0$}
         \State $\textbf{\textit{PANDA\_Mem_insert}}(k\_mer,1)$ 
         \Else  \State $New\_frq \gets \textbf{\textit{PANDA\_Add}}(k\_mer,1)$ \Comment{increment frq by 1}
                 \State $\textbf{\textit{PANDA\_Mem_insert}}(k\_mer,New\_frq)$ \Comment{insert into Hashmap again}
         \EndIf
    \EndFor
\State $\textbf{return}$ Hashmap
  \end{algorithmic}
  \end{minipage}}
\end{algorithm}

\begin{figure}[h]
\begin{center}
\begin{tabular}{c}
\includegraphics [width=1\linewidth]{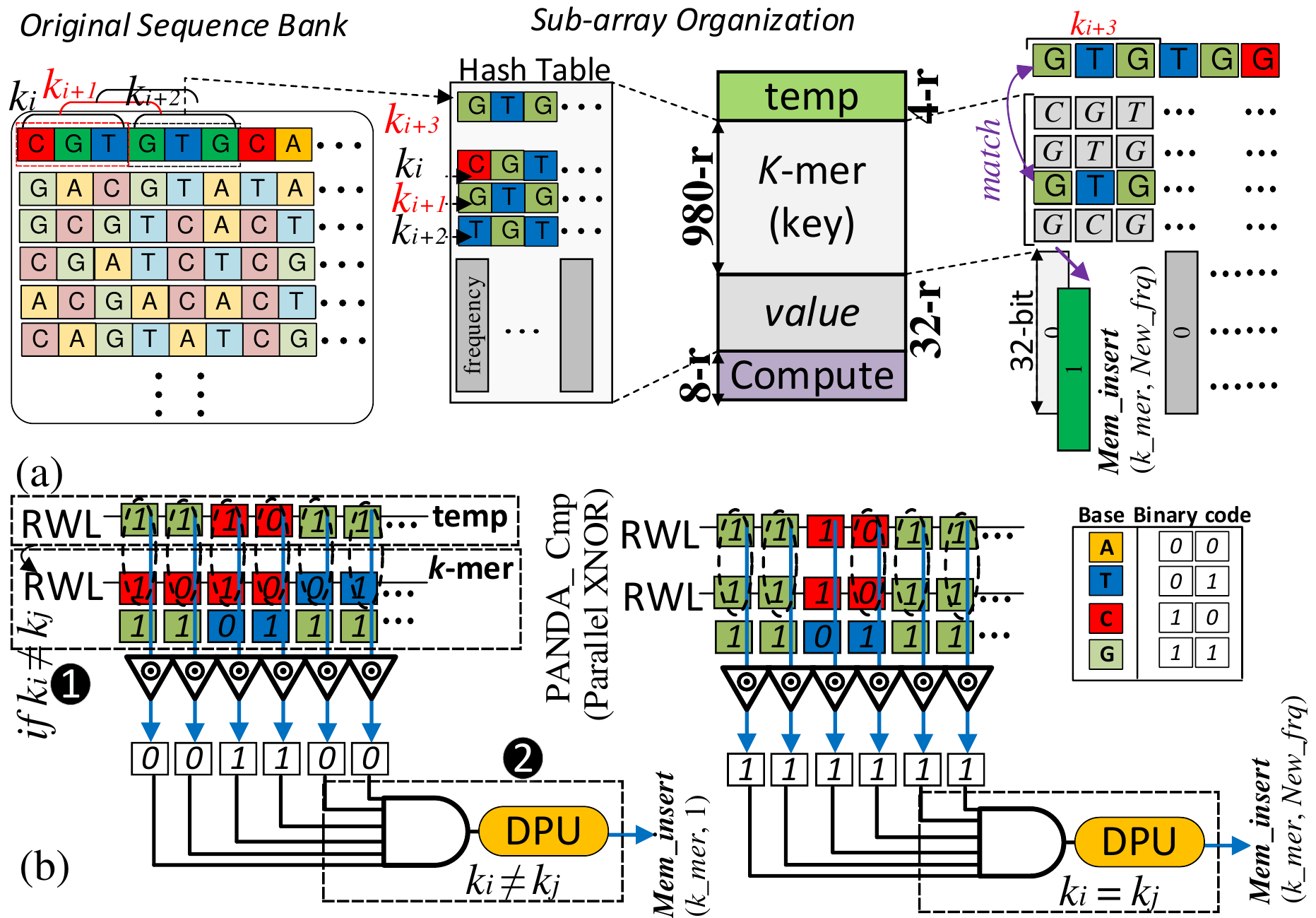}\\
\end{tabular} \vspace{-1.1em}
\caption{ (a) The proposed correlated data partitioning and mapping methodology for creating hash table, (b) Realization of parallel in-memory comparator (\textit{PANDA\_Cmp}) between \textit{k}-mers in a computational sub-array.}
\label{kmere1}
\end{center}\vspace{-1.2em}
\end{figure}

The proposed correlated partitioning and mapping methodology, as shown in Fig. \ref{kmere1}a, locally stores correlated regions of \textit{k}-mer (980 rows) vectors, where each row stores up to 128 bps (\textit{A,C,G,T} encoded by 2 bits) and value (32 rows) vectors in the same sub-array. For counting the frequencies of each distinct \textit{k}-mer, the ctrl first reads and parses the short reads from the original sequence bank to the specific sub-array. 
As depicted in Fig. \ref{kmere1}a, assuming \textit{S=CGTGTGCA} as the short read, the \textit{k}-mers- $k_{i}$-$k_{i+n}$ are extracted and written into the consecutive memory rows of \textit{k}-mer region. However, when a new query such as $k_{i+3}$ arrives (while $k_{i}$-$k_{i+2}$ are already in the memory), it will be first written to the temp region. A parallel in-memory comparison operation (\textit{PANDA\_Cmp}) will be performed between temp data and already-stored \textit{k}-mers. Fig. \ref{kmere1}b intuitively shows \textit{PANDA\_Cmp} procedure, where entire temp row can be compared with a previous \textit{k}-mer row in a single cycle. Then, a built-in ctrl's AND unit in DPU readily takes all the results to determine the next memory operation according to the algorithm. To increase the frequency of a specific \textit{k}-mer,  \textit{PANDA\_Add} is leveraged to perform in-memory addition without sending data to off-chip processor.

\subsection{Stage Two: Graph Construction}
\begin{figure}[t]
\begin{center}
\begin{tabular}{l}
\includegraphics [width=1.05\linewidth]{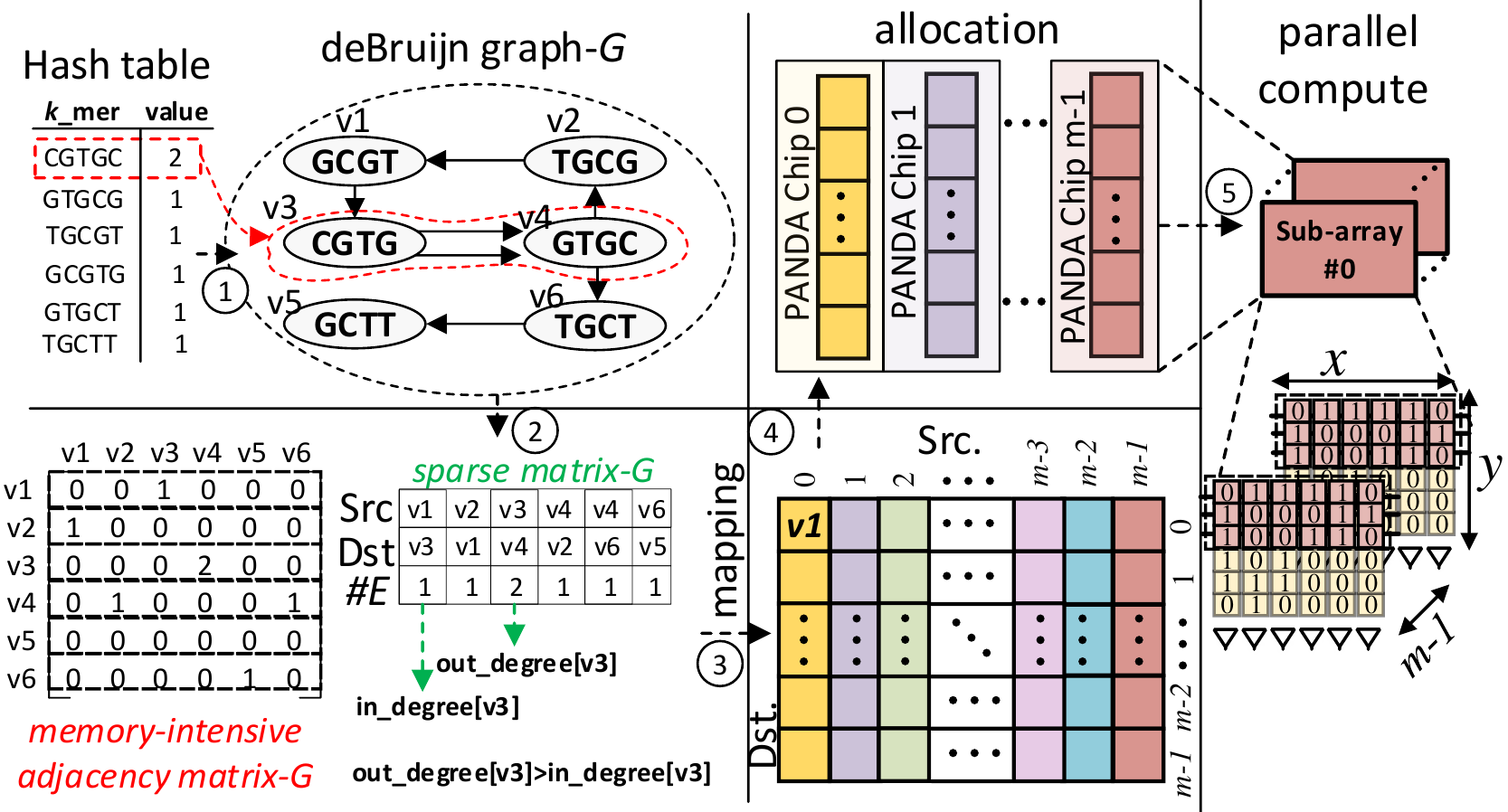}\\
\end{tabular} \vspace{-0.8em}
\caption{Graph construction with sparse matrix with partitioning,  allocation  and  parallel computation.}
\label{partition}
\end{center}\vspace{-0.5em}
\end{figure}

The next step is to construct and access a de Bruijn graph based on the Hash structure to rapidly lookup of a ‘‘value’’ associated with each \textit{k}-mer. For each entry (of length $k$) in the Hashmap, we will make two nodes, one with the prefix of length $k$-1 and other with the suffix of length $k$-1 (e.g. CGTGC $\to$ CGTG and GTGC), and connect an edge between them. For each Hash table entry with $n$ as the frequency, $n$ edges is then added between the two nodes. The de Bruijn graph $G$ for the sample Hash table in Fig. \ref{diag1} is constructed in Fig. \ref{partition} (step 1). Algorithm 2 shows the reconstructed de Bruijn procedure for \textit{PANDA} taking Hashmap data and \textit{k} as input returning matrix \textit{G}. For each key within Hash table, \textit{PANDA\_Mem\_insert} instruction creates an entry in $G$ for node1 and node2s.
Leveraging adjacency matrix representation for direct mapping of such humongous sparse graph into memory comes at a cost of significantly increased memory requirement and run time. The size of adjacency matrix will be V$\times$V for any graph with V nodes, where sparse matrix could be represented by a 3$\times$E matrix, where E is the total number of edges in the graph.
\textit{PANDA} utilizes sparse matrix representation shown in Fig. \ref{partition} (step 2) for mapping purpose. Each entry in the $3^{rd}$ row of the sparse matrix represents the number of connections between two nodes in $1^{st}$ and $2^{nd}$ rows.  
\begin{algorithm}[h]
  \caption{Procedure DeBruijn(Hashmap, k)}
  \label{algorithm1}
  \scalebox{0.67}{
    \begin{minipage}{1.5\linewidth}
  \begin{algorithmic}[1]
\Statex Step-1. Initialization:
\State G=[], Nodes\_List=[], i=1
\Statex Step-2. Sparse Graph Construction:
  \For{ $\forall  k\_mer\in Hashmap.keys()$, $i++$}
        \State $node\_1\gets k\_mer[0 : k-2]$ 
         \State $node\_2 \gets k\_mer[1 : k-1]$ 
          \State $\textbf{\textit{PANDA\_Mem_insert}}(G[1][i],node\_1)$ 
          \State $\textbf{\textit{PANDA\_Mem_insert}}(G[2][i],node\_2)$
          \State $\textbf{\textit{PANDA\_Mem_insert}}(G[3][i],Hashmap[k\_mer])$
 \EndFor
    \State $\textbf{return}$ {G}
  \end{algorithmic}
  \end{minipage}}
\end{algorithm}

To balance workloads of each \textit{PANDA}'s chip and maximize parallelism, we leverage interval-block partitioning method. We use hash-based approach \cite{dai2018graphh} by splitting the vertices into $M$ intervals and then divide edges into $M^2$ blocks as shown Fig. \ref{partition} (step 3: mapping). Then each block is allocated to a chip (step 4: allocation) and mapped to its sub-arrays. Having an $m$-vertex sub-graph with $N_s$ activated sub-arrays (size=$x\times y$), each sub-array can process $n$ vertices ($n \leq f|n\in N, f = min(x,y)$) (step 5: parallel computation). In this way, the number of  processing sub-arrays for an $N$-vertex sub-graph can be formulated as, $N_s= \left \lceil \frac{N}{f} \right \rceil$.

After graph construction, it is possible to perform a round of simplification on the sparse graph stored in \textit{PANDA} without loss of information to avoid fragmentation of the graph. As a matter of fact, the blocks are broken up each time a short read starts or ends leading to linear connected subgraphs \cite{zerbino2008velvet}. This fragmentation imposes longer execution time and larger memory space. The simplification process easily merges two nodes within memory if a node-A has only one out-going edge directed to node-B with only one in-going edge.

\subsection{Stage Three: Traversal for Euler Path}

The input of this stage will be a sparse representation of graph $G$. For traversing all the edges, we will use Fleury’s algorithm to find the Euler path of that graph (a path which traverses all edges of a graph). Basically, a directed graph has a Euler path if the in\_degree and out\_degree\footnote{The in\_degree[$i$] shows how many edges are coming into a vertex-$i$ and out\_degree[$i$] means how many out-going edges vertex-$i$ has.} of every vertex is same or, there are exactly two vertices which have |in\_degree - out\_degree|= 1. Finding the starting vertex is very important to generate the Eulerian path and we cannot consider any vertex as a starting vertex. The reconstructed PIM-friendly algorithm for finding the start vertex in graph-$G$ in shown in Algorithm 3. For each node, this stage deals with massive number of iteratively-used \textit{PANDA\_Add} to calculate the number of in\_degree, out\_degree and edge\_cnt (total number of edges). Moreover, in order to check the condition (|out\_degree = in\_degree|+ 1), parallel \textit{PANDA\_Cmp} operation is required. 

\begin{algorithm}[t]
  \caption{Procedure Find Start Vertex(G)}
  \label{algorithm2}
  \scalebox{0.67}{
    \begin{minipage}{1.5\linewidth}
  \begin{algorithmic}[1]
\Statex Step-1. Initialization:
    \State $start \gets 0$, $end \gets 0$
    \State $edge\_cnt \gets 0$  \Comment{For counting number of edges in $G$}
    \State $Len \gets size(G)$
\Statex Step-2. Find the start vertex:
\For{$n$ in Nodes}
        \State $in\_degree[i] \gets 0$ 
         \State $out\_degree[i] \gets 0$ 
\EndFor
\For{$n$ in Nodes}
     \For{$k:=$1 to $Len$}
           \If{$\textbf{\textit{PANDA_Cmp}}(G[1][k],n)$} \Comment{node n has an out-going edge}
           \State $out\_degree[n] \gets \textbf{\textit{PANDA\_Add}}(out\_degree[n], int(G[3][k])) $ 
           \State $in\_degree[int(G[2][k])] \gets \textbf{\textit{PANDA\_Add}}(in\_degree[int(G[2][k])], int(G[3][k])) $ 
           \State $edge\_cnt \gets \textbf{\textit{PANDA\_Add}}(edge\_cnt, int(G[3][k])) $ 
          \EndIf
    \EndFor
           \If{$\textbf{\textit{PANDA_Cmp}} (out\_degree[n],in\_degree[n]+1)$} 
           \State $start \gets n$ 
           \Else 
           \State $start \gets first\_node$
          \EndIf
      
\EndFor
 \State $\textbf{return}$ start \&  edge\_cnt \& out\_degree
  \end{algorithmic}
  \end{minipage}}
\end{algorithm}

After finding the start node, \textit{PANDA} has to traverse through the length of sparse matrix $G$ from the starting vertex and check two conditions for each edge and accordingly add qualified edges to the Eulerian Path. We show the reconstructed Fleury algorithm in Algorithm 4.  If an edge \textit{is not a bridge} and \textit{is not the last edge of the graph}, we will add $(start, v)$ in the Eulerian path and remove that edge.  $isValidNextEdge()$ function will check if the edge $(u,v)$ is valid to be included into our Euler path. If $v$ is the only adjacent vertex remaining for $u$, it means that, we have traversed all other adjacent vertices, so we will take this edge, otherwise we won’t. The second condition counts the number of reachable nodes from $u$ before and after removing the edge. If the number changes/decreases, it means that, the edge was a bridge (removing it will disconnect the graph into two parts). If it is a bridge, we cannot remove the edge from our Graph; otherwise we will remove the edge and add it into Euler path.

\begin{algorithm}[h]
  \caption{Procedure  Fleury(G, node, edge\_count, out\_degree)}
  \label{algorithm3}
  \scalebox{0.67}{
    \begin{minipage}{1.5\linewidth}
  \begin{algorithmic}[1]
    \For{$v$ := 0 to $N$} 
     \If{$G[1][k]== start$}
     \State $v \gets G[2][k]$
       \If{$isValidNextEdge(v)$}
        \State $\textbf{\textit{PANDA\_Mem_insert}}(v)$ \Comment{  add ($start$, $v$) in the Eulerian path}
         \State  $\textbf{\textit{PANDA\_Add}}(out\_degree[start], -1)$
         \State  $\textbf{\textit{PANDA\_Add}}(G[3][k], -1)$ \Comment{remove one edge from the graph}
         \State  $\textbf{\textit{PANDA\_Add}}(edge\_cnt, -1)$
     \EndIf
     \EndIf
  \State Fleury(G, v, edge\_count, out\_degree[]) \Comment{run Fleury again for the next node $v$}
       \EndFor
  \end{algorithmic}
  \end{minipage}}
\end{algorithm}

\begin{figure}[t]
\begin{center}
\begin{tabular}{l}
\includegraphics [width=0.99\linewidth]{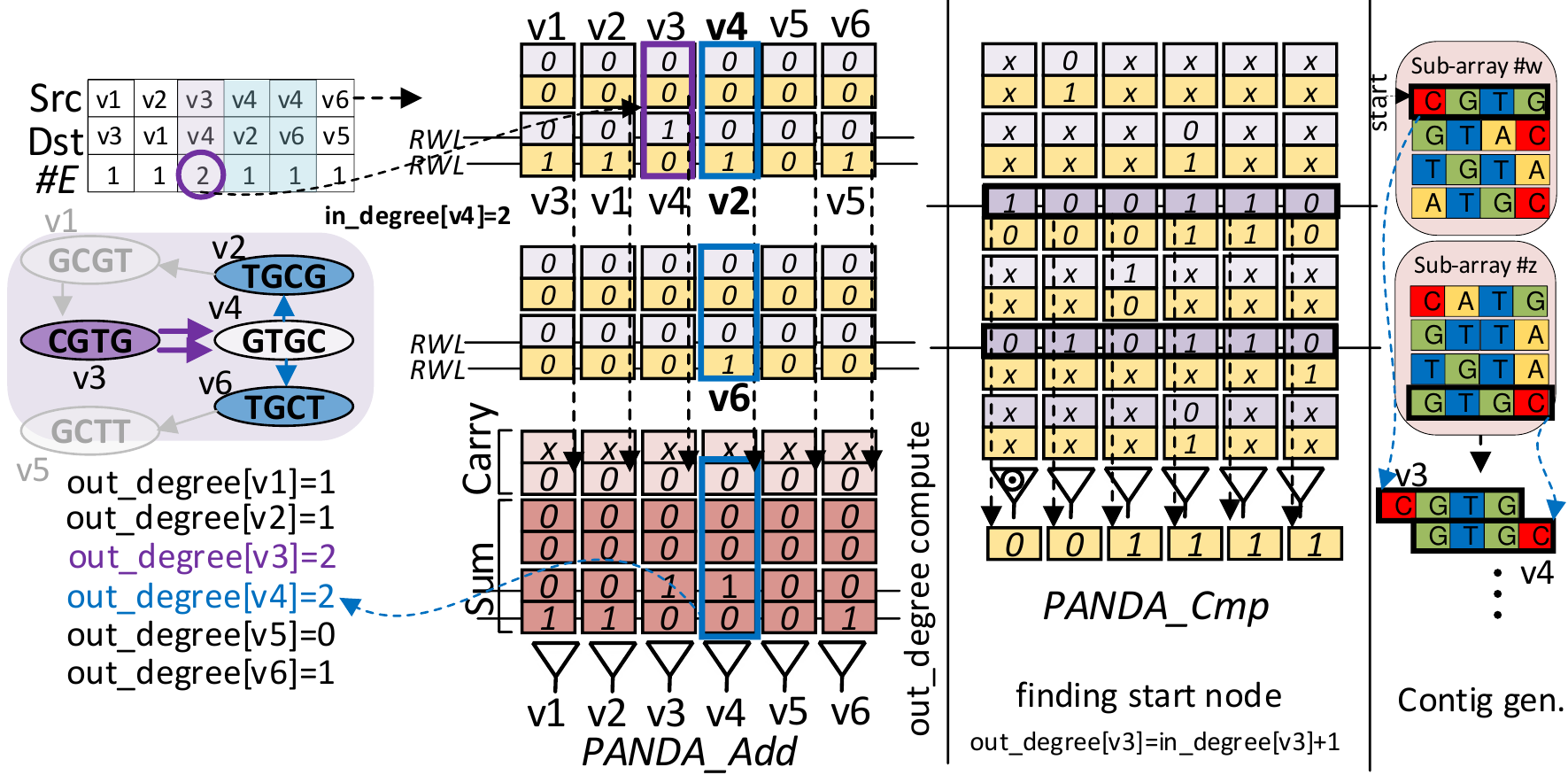}\\
\end{tabular} \vspace{-1em}
\caption{\textit{PANDA} in-memory addition and comparison scheme for finding the start vertex.}
\label{partition2}
\end{center}\vspace{-1.8em}
\end{figure}

In the interest of space, we show out\_/in\_degree and edge\_cnt mapping and computation in the \textit{PANDA} platform in Fig. \ref{partition2}, which basically sums up all the entries of a particular node \textit{i} of valid links connected to a vertex to find the start vertex. As can be seen, we use the sparse matrix representation to store the matrix-$G$. In our mapping technique, each column is assigned to a distinct source vertex in the graph and then filled out with the number of edges (\#E) only linked to existing destination vertices in a vertical fashion. Therefore, we do not assign destination vertices to the memory rows as in direct adjacency matrix mapping. 
Here, we consider a 4-bit representation for the simplicity. For example, v4 has out-going edges to v2 and v6 that are stored vertically in a sub-array. \textit{PANDA} could perform parallel in-memory addition to calculate the total number of out\_ degree for all nodes in parallel. For this task, two rows in the sub-array are initialized to zero as Carry reserved rows such that they can be selected along with two operands (here v4$\to$v2 data (0001) and v4$\to$v6 data (0001)) to perform parallel in-memory addition. 
To perform parallel addition operation and generate initial Carry and Sum bits, \textit{PANDA} takes every three rows to perform a parallel in-memory addition. The results are written back to the memory reserved space (Resv.). Then, next step only deals with multi-bit addition of resultant data starting bit-by-bit from the LSBs of the two words continuing towards MSBs.
Then \textit{PANDA} is able to perform comparison between number of out\_degree and in\_degree for each node in parallel to determine the start node. After finding the start node as shown in Fig. \ref{partition2}, contig. generation can be readily accomplished through finding the Eulerian path and putting together each vertex data from different sub-arrays.

\section{Performance Estimation}

\subsection{Setup}
\begin{figure}[h]
\begin{center}
\begin{tabular}{l}
\includegraphics [width=0.99\linewidth]{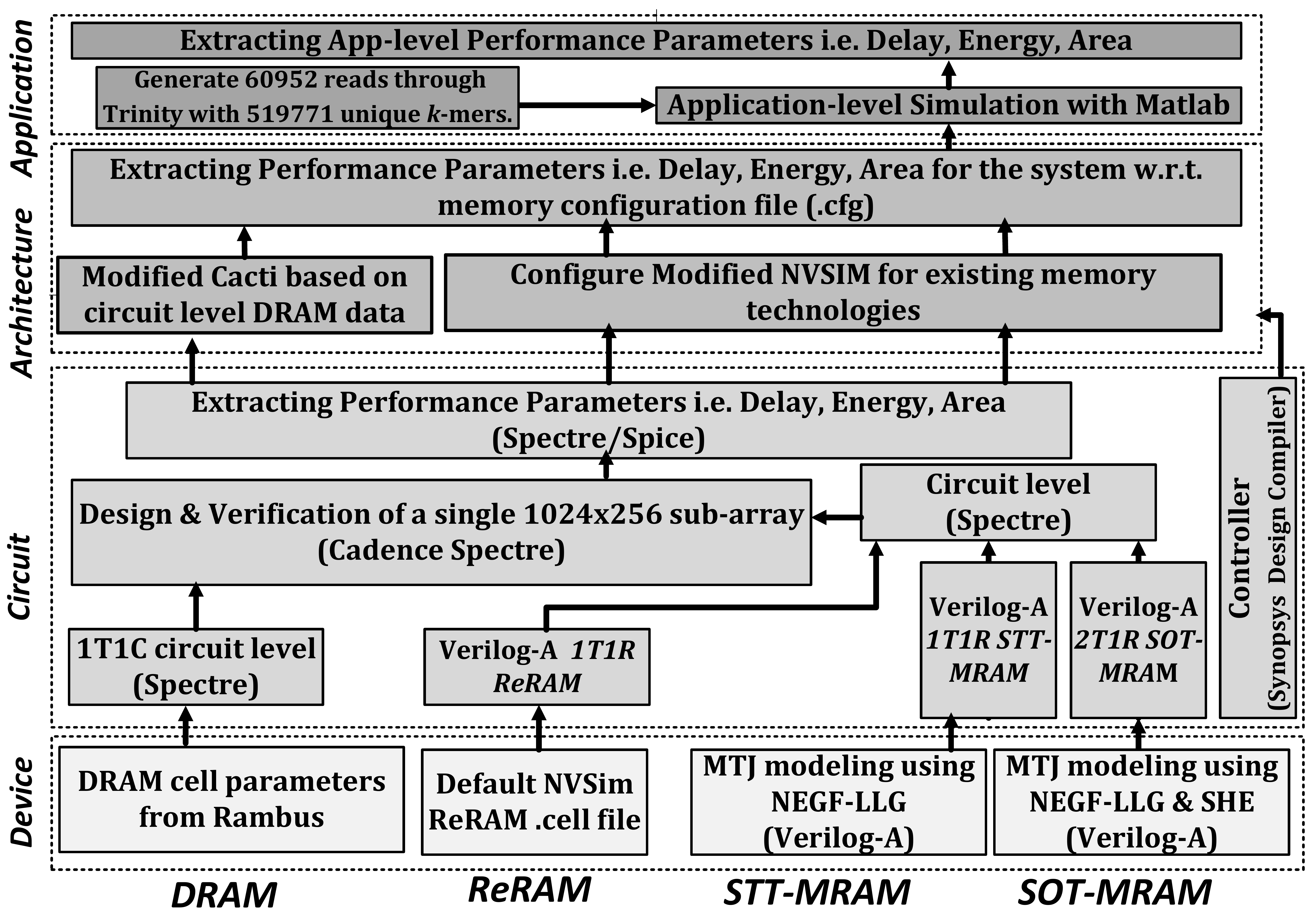}\\
\end{tabular} \vspace{-1em}
\caption{Evaluation framework developed for processing-in-memory platforms.}
\label{FC}
\end{center}\vspace{-1em}
\end{figure}

\textit{\textbf{Accelerator:}}
To the best of our knowledge, this work is the first to explore the performance of a PIM platform for genome assembly problem, therefore, we have to create the evaluation test bed from scratch to have an impartial comparison with both von-Neumann and non-von-Neumann architectures. We configure the \textit{PANDA}'s computational memory sub-array with 1024 rows and 256 columns, 4$\times$4 memory matrix (with 1/1 as row/column activation) per bank organized in H-tree routing manner, 16$\times$16 banks (with 1/1 as row/column activation) in each memory chip. For comparison, we consider five computing platforms: 1) A general purpose processor (GPP): a Quad Core Intel Core i7-7700 CPU @ 3.60GHz processor with 8192MB DIMM DDR4 1600MHz RAM and 8192KB Cache; 2) A processing-in-STT-MRAM platform capable of performing bulk bit-wise operations \cite{jain2017computing}; 3) A recently developed processing-in-SOT-MRAM platform for DNA sequence alignment optimized to perform comparison-intensive operations \cite{angizi2019aligns}; 4) A processing-in-ReRAM accelerator designed for accelerating bulk bit-wise operations \cite{imani2017mpim}; 5) A processing-in-DRAM accelerator based on Ambit \cite{seshadri2017ambit} working with triple row activation mechanism to implement various functions. The detailed evaluation framework developed for PIM platforms is shown in Fig. \ref{FC}. 
All PIM platforms have an identical physical memory configuration as \textit{PANDA}. Additionally, we developed a similar cross-layer simulation framework starting from device-level simulation all the way to circuit- and architectural level as explained for \textit{PANDA} in Section II.D. The results of the architecture evaluation of all PIM platforms were then fed to a high-level in-house simulator developed in Matlab to perform each genome assembly stage based on our customized and PIM-friendly algorithm and estimate the overall performance. It is noteworthy that DPU was developed in HDL and the performance results was extracted with synopsys design compiler \cite{DC} and fed to the developed NVSim library for each PIM platform.

To evaluate the CPU performance, we use Trinity-v2.8.5 \cite{grabherr2011full} which was shown to be sensitive and efficient in recovering full-length transcripts. Trinity constructs de Bruijn graph from short-read sequences and employs an enumeration algorithm to score all branches, and keeps possible ones as isoforms/transcripts. 

\textit{\textbf{Experiment:}} In our experiment, we create 60952 short reads through Trinity sample genome bank with 519771 unique \textit{k}-mers. We initially set the \textit{k}-mer length, $\textit{k}$, to default 25, and then change it to 22, 27, and 32 as typical values for most genome assemblers. To clarify, the CPU executes the \textit{Inchworm}, \textit{Chrysalis}, and \textit{Butterfly} steps in Trinity, while PIM platforms run three main procedures in genome assembly shown in Fig. \ref{diag} i.e. Hashmap, DeBruijn, and Traverse for under-test PIM platforms. We compare Trinity's power consumption and execution time to that of other PIM assemblers by several measures. To have a fair comparison with such a comprehensive assembler (that performs full genome assembly task with scaffolding step), we penalized the PIM platforms with $\sim$25\% excessive time and power. We believe this could provide a more realistic comparison with a von-Neumann architecture-based assembler.   

\vspace{-0.8em}

\begin{figure}[b]
\begin{center}
\begin{tabular}{c}
\includegraphics [width=0.99\linewidth]{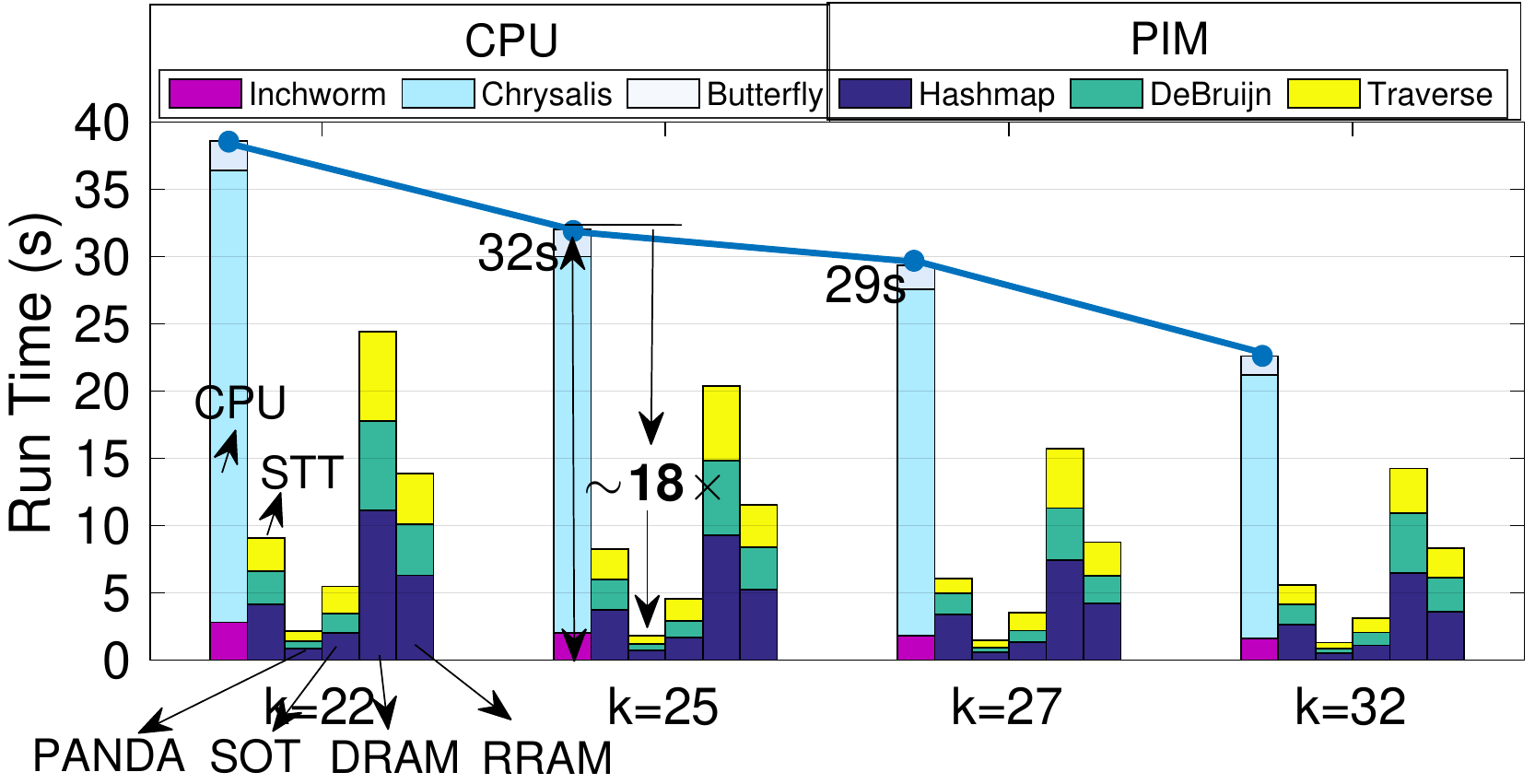}\\
 \end{tabular}\vspace{-0.7em}
\caption{The breakdown of run time for under-test platforms running different \textit{k}-mer-length genome assembly task. In each bar group from left to right: CPU, processing-in-STT-MRAM \cite{jain2017computing}, \textit{PANDA}, processing-in-SOT-MRAM \cite{angizi2019aligns}, processing-in-DRAM \cite{seshadri2017ambit}, and processing-in-RRAM \cite{imani2017mpim}.}
\label{perf}
\end{center}\vspace{-1.5em}
\end{figure}

\begin{figure}[t]
\begin{center}
\begin{tabular}{c}
\includegraphics  [width=0.98\linewidth]{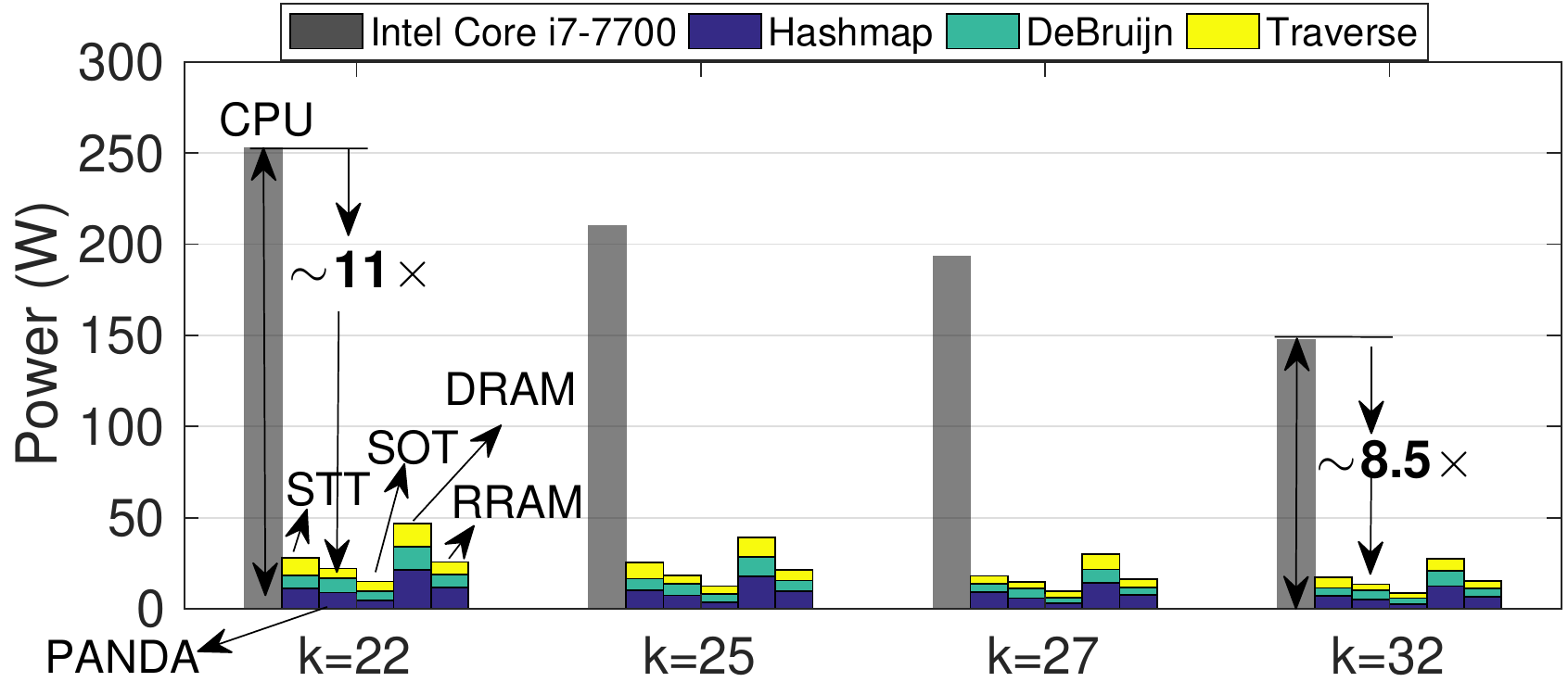}\\
 \end{tabular}\vspace{-0.5em}
\caption{The breakdown of power consumption for PIM platforms running different \textit{k}-mer-length genome assembly task compared to CPU. In each bar group from left to right: CPU, processing-in-STT-MRAM \cite{jain2017computing}, \textit{PANDA}, processing-in-SOT-MRAM \cite{angizi2019aligns}, processing-in-DRAM \cite{seshadri2017ambit}, and processing-in-RRAM \cite{imani2017mpim}.}
\label{power}
\end{center}\vspace{-2.1em}
\end{figure}

\subsection{Run Time}
The execution time of genome assembly task for different platforms is reported in Fig. \ref{perf}. For \textit{k}=25, the CPU platform executes the \textit{Inchworm}, \textit{Chrysalis}, and \textit{Butterfly} steps \cite{grabherr2011full} of Trinity in $\sim$32s, where Chrysalis for clustering the contigs and constructing complete de Bruijn graph takes the largest fraction of the run time (28s) as expected. However, the comparison operation-intensive Hashmap procedure for \textit{k}-mer analysis takes the largest fraction of execution time in all PIM platforms (over 40\% of total run time).
Larger \textit{k}-mer length typically diminishes the de Bruijn graph connectivity by simultaneously reducing the number of ambiguous repeats in the graph and chance of overlap between two reads. That is why run time for all platforms reduces with increase of \textit{k}-mer length.

We can observe that PIM platforms reduce the run time remarkably compared to the CPU. As shown, \textit{PANDA} reduces the run time by $\sim$18$\times$ compared to the CPU platform for \textit{k}=25 (18.8$\times$ on average over 4 different \textit{k}-mer lengths). 
The \textit{PANDA} platform essentially accelerates the graph construction and traversal stages by $\sim$21.5$\times$ compared with CPU platform. Now, by increasing the \textit{k}-length to 32, the higher speed-up is even achievable.
Compared with counterpart PIM platforms, our X(N)OR-friendly design reduces the run time on average by 4.2$\times$, 2.5$\times$, compared to STT-PIM \cite{jain2017computing}, and SOT-PIM \cite{angizi2019aligns} platforms as the fastest counterparts, respectively. This comes from the fact that under-test PIM platforms require multi-cycle operations to implement addition operation. Besides, the SOT-based device intrinsically shows higher write speed compared to STT devices. Compared to DRAM and RRAM platforms, \textit{PANDA} achieves on average 10.9$\times$ and 6$\times$ speed-up for various length $k$-mer processing. It is worth pointing out that the processing-in-DRAM platforms possess a destructive computing operation and require multiple memory cycle to copy the operands to particular rows before computation. As for Ambit \cite{seshadri2017ambit}, 7 memory cycles are needed to implement in-memory-X(N)OR function.

\vspace{-1em}

\subsection{Power Consumption}
We estimated the power consumption of different PIM platforms for running different length \textit{k}-mers compared to the CPU platform as shown in Fig. \ref{power}. Based on our results, a significant reduction in power consumption can be reported for all under-test PIM platforms compared with the CPU. The breakdown of energy consumption is also shown for the PIM platforms, however this couldn't be accurately achieved for the CPU and overall power consumption is reported. In our experiment, processing-in-SOT-MRAM design \cite{angizi2019aligns} achieves the smallest power consumption (on average) to run the three main procedures, as compared with the CPU and other PIM platforms. The \textit{PANDA} platform stands as the second most power-efficient design. This is mainly due to the three-SA based bit-line computing scheme in \textit{PANDA} compared with two-SA per bit-line technique in the counterpart design. While the proposed scheme brings more speed-up compared with the design in \cite{angizi2019aligns}, it requires relatively more power. The \textit{PANDA} reduces the power consumption by $\sim$9.2$\times$ on average compared with the CPU platform over different length \textit{k}-mers. Besides, it reduces the power consumption by $\sim$18\% compared with STT-MRAM \cite{jain2017computing} platform. The main reason behind this improvement is more efficient addition operation in \textit{PANDA}. Addition operation requires additional memory cycles in the STT-MRAM \cite{jain2017computing} platform to save carry bit back to the memory and use it again for the computation of next bits. Compared to DRAM and RRAM platforms, \textit{PANDA} obtains on average 2.11$\times$ and 55\% power reduction for various length $k$-mer processing.

\subsection{Speed-up/Power-Efficiency Trade-off}
\begin{figure}[t]
\begin{center}
\begin{tabular}{c}
\includegraphics [width=0.95\linewidth]{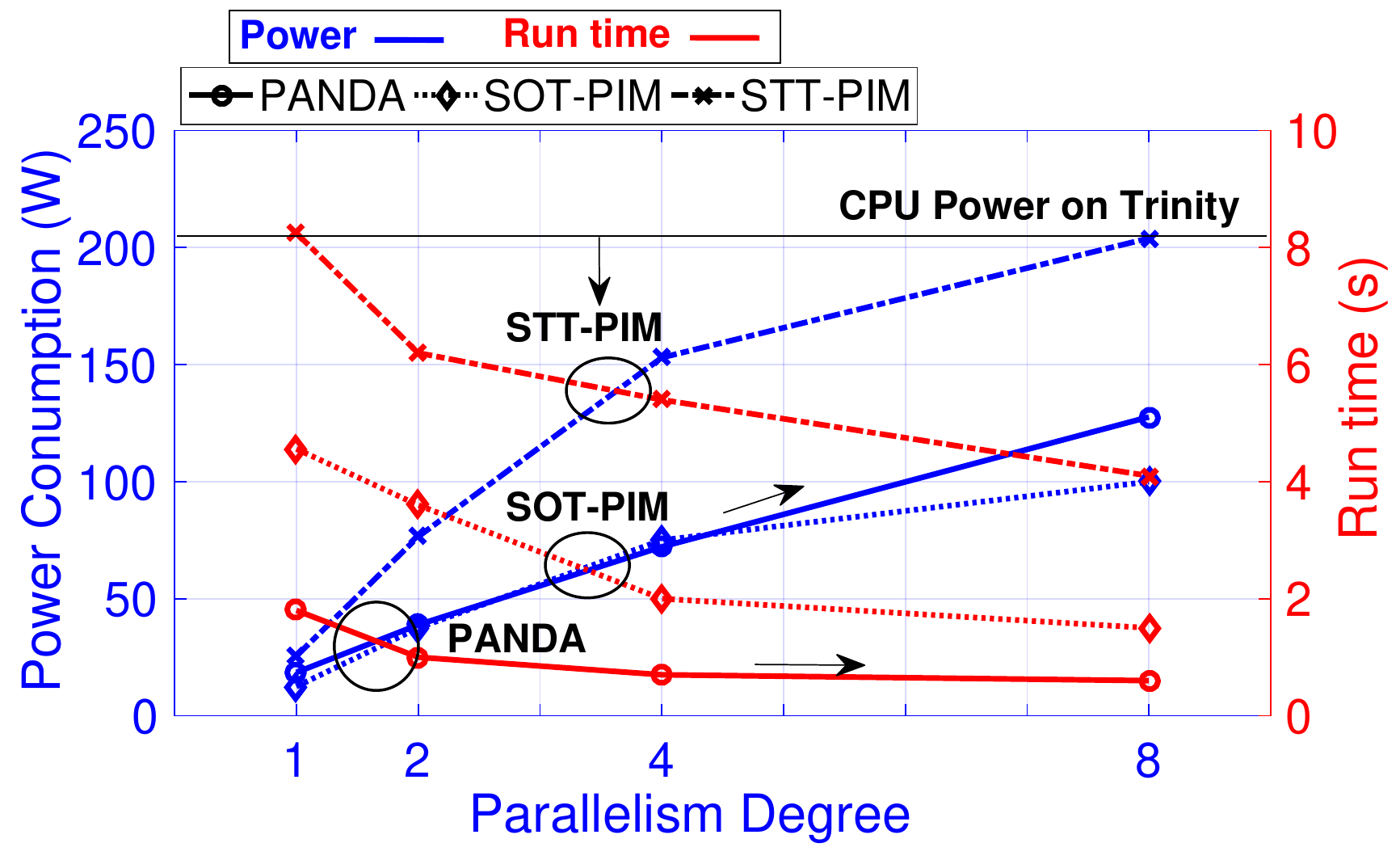}\\
 \end{tabular}\vspace{-1.2em}
\caption{Trade-off between power consumption and run-time w.r.t parallelism degree in \textit{k}=25.}
\label{pd}
\end{center}\vspace{-1.6em}
\end{figure}

We investigate the power-efficiency and speed-up of three best under-test PIM platforms, based on the run time and power consumption results in the previous subsections, by tuning the number of active sub-arrays ($N_s$) associated with the 
comparison and addition operations. A parallelism degree ($P_d$) can be then defined as the number of replicated sub-arrays to boost the performance of the PIM platforms through parallel processing as shown in prior works \cite{angizi2019aligns,li2016pinatubo}. For example, when $P_d$ is set to 2, two parallel sub-arrays are undertaken to process the in-memory operations, simultaneously. We expect such parallelism to improve the performance of genome assembly at the cost of sacrificing the power consumption and area.
Fig. \ref{pd} plots the existing trade-off between run time and power consumption vs. $P_d$ for \textit{k}= 25. The estimated CPU power budget required to execute Trinity is also shown. It can be seen that for all platforms the run time reduces by increasing the parallelism. For example for \textit{PANDA} platform in an extreme case, increasing $P_d$ from 1 to 8 increases the power consumption from $\sim$19W to 128W ($\sim$7$\times$) and reduces the execution time by a factor of 3, which might not be a favorable case. Therefore, a user can meticulously tailor the \textit{PANDA} performance to meet the system/application constraints. Here, we show the optimum theoretical performance of \textit{PANDA} and other PIM platforms by pinpointing the intersection between power and run time curves in Fig. \ref{pd}. We observe that \textit{PANDA} achieves the smallest run time and power consumption task with a $P_d \sim$2 compared with the others.
\vspace{-0.9em}

\subsection{Memory Wall Challenge}
\begin{figure}[h]
\begin{center}
\begin{tabular}{c}
\includegraphics [width=0.99\linewidth]{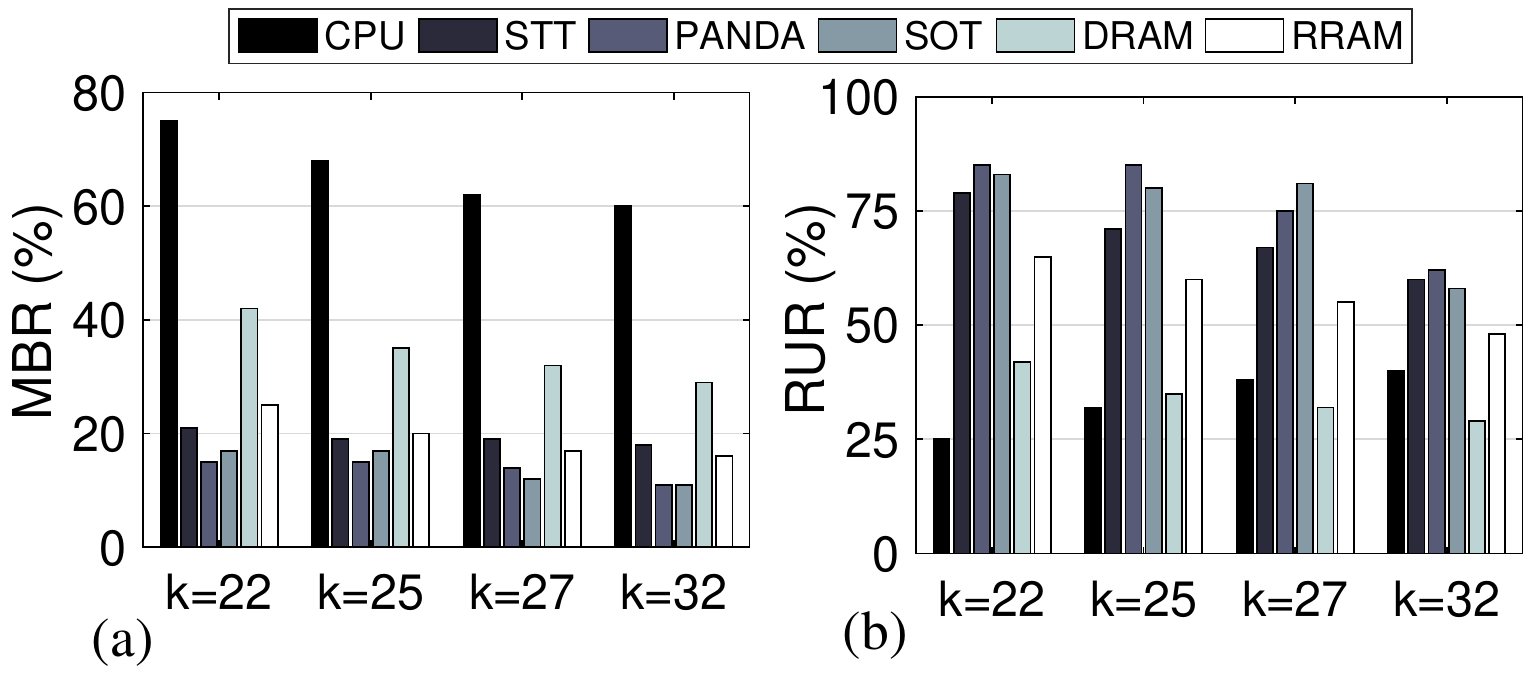}\vspace{-0.7em}
\end{tabular}
\caption{(a) The memory bottleneck ratio and (b) resource utilization ratio for CPU and three under-test PIM platforms for running genome assembly task.}
\label{memorywall}\vspace{-1.7em}
\end{center}
\end{figure}

The power-efficiency and speed-up of PIM platforms against the von-Neumann architecture-based CPU was discussed in prior subsections. Here, we further explore the reasons behind the numbers reported by considering two new measures i.e. Memory Bottleneck Ratio (MBR) and Resource Utilization Ratio (RUR). We define MBR as the time fraction needed for data transfer from/to on-chip or off-chip, when computation has to wait for data i.e. memory wall happens. We also define RUR as the time fraction in which the computation resources are loaded with data. The memory wall is considered as the main bottleneck that brings large power consumption and lengthen execution time in CPU.
The MBR is reported in Fig. \ref{memorywall}a. The peak throughput for each design in four distinct \textit{k}-mer lengths is taken into account for performing the evaluation. This evaluation mainly considers the number of memory access.
As shown, the \textit{PANDA} uses less than $\sim$17\% time for data transfer due to the PIM acceleration schemes, while CPU's MBR increases to 65\% when $k$=25. Besides, we observe that all the other PIM platforms except DRAM also spend less than $\sim$17\% time for data communication.
The smaller MBR can be translated as the higher RUR for the accelerators plotted in Fig. \ref{memorywall}b. The less MBR can be understood as a higher RUR. We can see that with up to $\sim$ 82\%, \textit{PANDA} achieves the highest RUR. Taking everything into account, PIM acceleration schemes offer a high utilization ratio ($>$60\% excluding DRAM) confirming the conclusion drawn in Fig. \ref{memorywall}a. The memory wall evaluation shows the efficiency of the \textit{PANDA} platform for solving memory wall challenge. \vspace{-1em}

\section{Conclusion}
In this paper, we presented \textit{PANDA} as a new processing-in-SOT-MRAM platform to accelerate the comparison/ addition-extensive genome assembly application using PIM-friendly operations. We developed \textit{PANDA} based on a set of new circuit-level schemes to realize a data-parallel computational core for genome assembly. The platform is configured with a novel data partitioning and mapping technique that provides local storage and processing to fully utilize our customized algorithm-level's parallelism. The cross-layer simulation results demonstrate that \textit{PANDA} reduces the execution time and power respectively by $\sim$18$\times$ and $\sim$11$\times$ compared with the CPU. Besides, speed-ups of up-to 2-4$\times$ can be obtained over recent processing-in-MRAM platforms to perform the similar task. 

\bibliographystyle{IEEEtran}
\scriptsize
\bibliography{IEEEabrv,./reference1}

\begin{IEEEbiography}[{\includegraphics[width=1in,height=1.25in,clip,keepaspectratio]{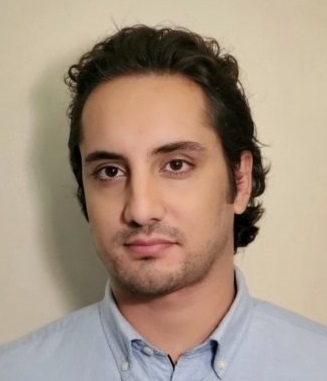}}]{Shaahin Angizi} is currently a Ph.D. candidate in Electrical Engineering at School of Electrical, Computer and Energy Engineering, Arizona State University, Tempe, AZ, USA. His primary research interests include ultra-low power in-memory computing based on volatile \& non-volatile memories, brain inspired (neuromorphic) computing, and accelerator design for deep neural network and bioinformatics. He is the recipient of Best Poster Award of Ph.D. Forum at IEEE/ACM DAC in 2018, two Best Paper Awards of IEEE ISVLSI in 2017 and 2018, and Best Paper Award of ACM GLSVLSI in 2019. He is a student member of IEEE.
\end{IEEEbiography}

\begin{IEEEbiography}[{\includegraphics[width=1in,height=1.25in,clip,keepaspectratio]{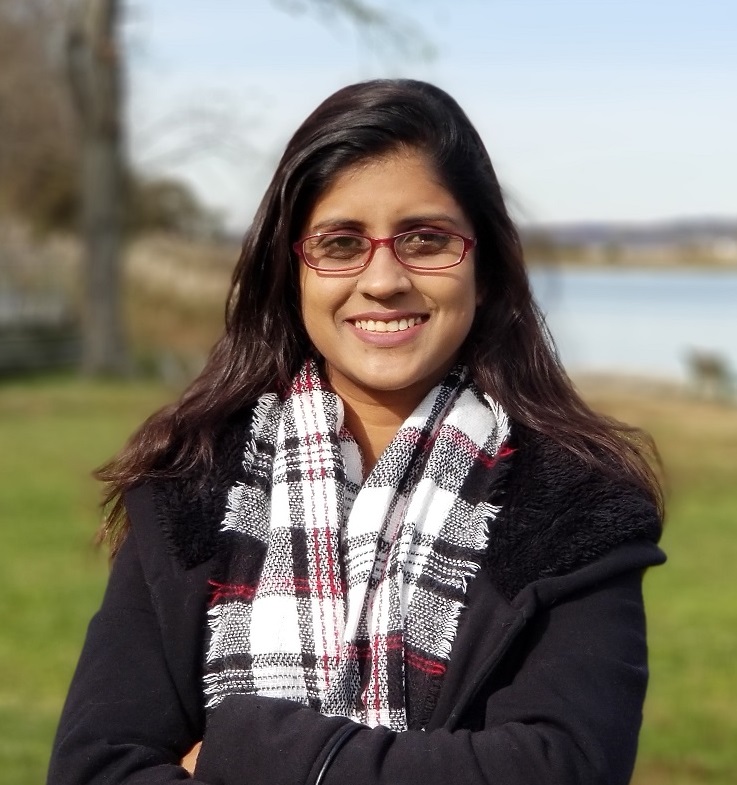}}]{Naima Ahmed Fahmi} is a Ph.D. student in Computer Science at University of Central Florida, Orlando, Florida. She is working under the supervision of Dr. Wei Zhang, Assistant Professor in Computer Science, University of Central Florida. Naima's research focus is to apply Machine Learning methods and algorithms to extract useful information from large scale genomic data. Currently, she is working with Cancer cell-lines and real patient's samples to predict and prognosis molecular markers for the disease.
\end{IEEEbiography}

\begin{IEEEbiography}[{\includegraphics[width=1in,height=1.25in,clip,keepaspectratio]{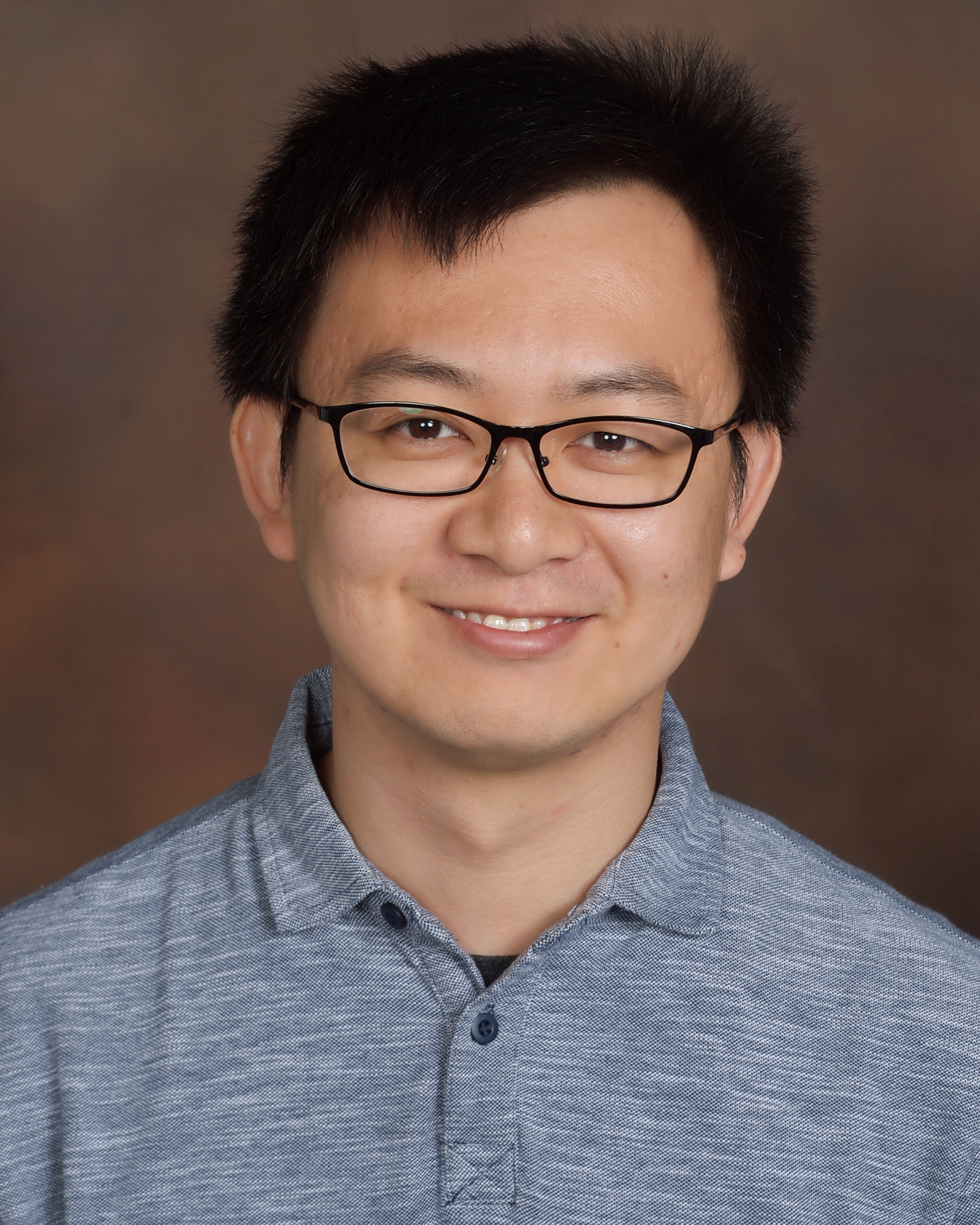}}]{Wei Zhang} received the MS and PhD degrees in computer science from University of Minnesota Twin Cities in 2011 and 2015, respectively. He joined the Department of Computer Science, University of Central Florida, Orlando, Florida, as an assistant professor in 2017. His primary research interest lies at the interaction of computational biology and machine learning, and has been focusing on graph-based learning models for biomarker selection and cancer outcome prediction. His other research interests include cancer transcript variants and drug sensitivity prediction. He received NSF CRII Award in 2018.
\end{IEEEbiography}

\begin{IEEEbiography}[{\includegraphics[width=1in,height=1.25in,clip,keepaspectratio]{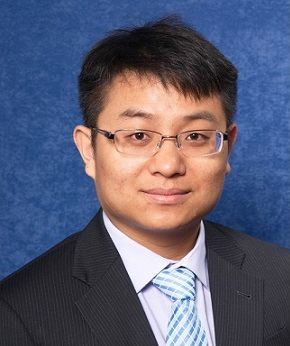}}]{Deliang Fan} is currently an Assistant Professor in the School of Electrical, Computer and Energy Engineering, Arizona State University, Tempe, AZ, USA. Before joining ASU in 2019, he was an assistant professor in Department of ECE at University of Central Florida, Orlando, FL, USA. He received his M.S. and Ph.D. degrees, under the supervision of Prof. Kaushik Roy, in ECE from Purdue University, West Lafayette, IN, USA, in 2012 and 2015, respectively. His primary research interests include Energy Efficient and High Performance Big Data Processing-In-Memory Circuit, Architecture and Algorithm, with applications in Deep Neural Network, Data Encryption, Graph Processing and Bioinformatics Acceleration-in-Memory system; Brain-inspired (Neuromorphic) and Boolean Computing Using Emerging Nanoscale Devices like Spintronics and Memristors; security of AI system. He has authored and co-authored 100+ peer-reviewed international journal/conference papers. He served as the TPC member of DAC, ICCAD, DATE, GLSVLSI, ISVLSI, ASP-DAC, ISQED, etc. He also served as the area technical chair of GLSVLSI, ISQED, etc. 
\end{IEEEbiography}

\end{document}